\documentclass[twocolumn]{aastex701}

\usepackage{amsmath}
\usepackage[utf8]{inputenc}
\usepackage{graphicx}
\usepackage{multirow}

\usepackage{lipsum}

\newcommand{\vect}[1]{\boldsymbol{#1}}

\newcommand{\dif}{\mathrm{d}}
\newcommand{\e}{\mathrm{e}}
\newcommand{\yr}{\,\mathrm{yr}}
\newcommand{\Myr}{\,\mathrm{Myr}}
\newcommand{\Gyr}{\,\mathrm{Gyr}}
\newcommand{\AU}{\,\mathrm{AU}}
\newcommand{\pc}{\,\mathrm{pc}}

\newcommand{\MSol}{M_{\odot}}
\newcommand{\RSol}{R_{\odot}}

\newcommand{\kms}{\mathrm{km \, s^{-1}}}

\newcommand{\lhat}{\vect{\hat{l}}}

\newcommand{\cmc}{{\tt CMC}}
\newcommand{\cosmic}{{\tt COSMIC}}
\newcommand{\catalog}{{\tt CMC Cluster Catalog}}

\newcommand{\ciera}{\affiliation{Center for Interdisciplinary Exploration and Research in Astrophysics (CIERA), Northwestern University, 1800 Sherman Ave., Evanston, IL 60201, USA}}
\newcommand{\nupa}{\affiliation{Department of Physics and Astronomy, Northwestern University, 2145 Sheridan Rd., Evanston, IL 60208, USA}}
\newcommand{\ucsd}{\affiliation{Department of Astronomy and Astrophysics, University of California, San Diego, La Jolla, CA 92093, USA}}

\shorttitle{BBHs from CMC Models with Realistic Binaries}
\shortauthors{C.\ E.\ O'Connor et al.}

\allowdisplaybreaks

\defcitealias{Kremer2020catalog}{K20}
\defcitealias{Offner2023binaries}{O23}

\begin{document}

\title{Black hole mergers from dense star clusters with realistic binary populations}

\author[0000-0003-3987-3776, sname='O\'Connor', gname='Christopher E.']{Christopher E.\ O'Connor}
\ciera
\email[show]{chrisoc@northwestern.edu}

\author[0000-0002-4086-3180, sname='Kremer', gname='Kyle']{Kyle Kremer}
\ucsd
\email{kykremer@ucsd.edu}

\author[0009-0007-6748-4627]{Saloni Agrawal}
\ucsd
\email{saagrawal@ucsd.edu}

\author[0000-0002-0933-6438]{Elena Gonz\'{a}lez Prieto}
\ciera
\nupa
\email{elena.prieto@northwestern.edu}

\author[0000-0003-4412-2176]{Fulya K{\i}ro\u{g}lu}
\ciera
\email{fulya.kiroglu@northwestern.edu}

\author[0000-0001-5285-4735]{Maia A.\ S.\ Martinez}
\ciera
\nupa
\email{miguel.martinez@northwestern.edu}

\author[0000-0001-9582-881X]{Claire S.\ Ye}
\affiliation{Canadian Institute for Theoretical Astrophysics, University of Toronto, 60 St.\ George St., Toronto, ON M5S 3H8, Canada}
\email{claireshiye@cita.utoronto.ca}

\author[0000-0002-7132-418X,gname='Frederic A.',sname='Rasio']{Frederic A.\ Rasio}
\ciera
\nupa
\email{rasio@northwestern.edu}

\correspondingauthor{Chris O'Connor}

\begin{abstract}
    We present a suite of $24$ full-lifetime simulations of dense star clusters with the \texttt{Cluster Monte Carlo} (\cmc) code, 
    featuring updated input physics and a realistic distribution of initial binary systems. 
    The latter encompasses a mass-dependent binary fraction, period distribution, and eccentricity distribution 
    based on observations of well-studied stellar populations in the Solar neighborhood and nearby star-forming regions. 
    We predict the cosmic rate, masses, and spins of binary black hole (BBH) mergers formed through dynamical assembly, primordial binary evolution, and hierarchical mergers within dense clusters. 
    As with previous model grids with fewer binaries, dynamically assembled first-generation (1G) mergers dominate the rate of cluster-derived mergers, 
    and the total merger rate is consistent with that inferred from LIGO--Virgo--KAGRA observations as of GWTC-5.0. 
    Our models naturally reproduce key features of the inferred BBH population, including 
    the broken-power-law behavior of the primary BH mass spectrum for $m_{1} \gtrsim 20 \MSol$, 
    the shallower (steeper) slope of the secondary mass spectrum relative to the primary for $m_{2} \lesssim 10 \MSol$ ($m_{2} \gtrsim 30 \MSol$), 
    and the shape of the mass-ratio distribution in the low- and high-mass domains. 
    We predict broad distributions of the spin parameters $\chi_{\rm eff}$ and $\chi_{\rm p}$, consistent with previous studies of dynamical assembly in clusters.  
    The merger rate from primordial binary systems within clusters is a small fraction of the total; 
    however, their merger products are frequently involved in subsequent hierarchical mergers, 
    with the result that the hierarchical merger rate evolves more steeply than the 1G dynamical merger rate with redshift. 
\end{abstract}


\section{Introduction} \label{s:Intro}

A decade after the first detection of gravitational waves (GWs) emitted from the inspiral and merger of two stellar-mass black holes (BHs), 
the astrophysical origin of binary black holes (BBHs) remains an open question. 
Several BBH formation channels have been proposed to operate in a variety of astrophysical environments 
\citep[e.g.,][]{BetheBrown1998, PortegiesZwart2001, Hurley+2002, McKernan+2012, Rodriguez2015, Rodriguez+2016, Bird+2016, MandeldeMink2016, deMinkMandel2016, Hoang+2018, MandelBroekgaarden2022}.  
With GW signals detected by the LIGO--Virgo--KAGRA (LVK) collaboration numbered in the hundreds as of the release of GWTC-5.0 \citep{LVK2026_gwtc5}, 
detailed analyses of BBH demographics, capable of disentangling the contributions of different formation environments, are becoming feasible \citep[e.g.,][]{Zevin+2021, LVK2025_populations, LVK2026_populations, Farah+2026, Ray+2026}. 

Star formation occurs predominantly in clusters \citep[e.g.,][]{LadaLada2003}, and the massive progenitor stars of stellar-mass BHs can live their whole lives within their dynamically active birth environments. 
Dense star clusters, from young massive clusters (YMCs) to old globular clusters (GCs), are therefore expected to be important sites of BH formation and mergers  
\citep[e.g.,][]{PortegiesZwart2001, Morscher2013bhdynamics, Morscher2015bhdynamics, Rodriguez2015, Chatterjee2017, Hong2018bbhmergers, Hong+2020, Rodriguez+2019, Kremer2020massgap, AntoniniGieles2020bbhpopsynth, Banerjee2021bbhclusters, Weatherford2021cmcimf}. 
Large grids of full-lifetime cluster models, surveying a wide range of masses, densities, and metallicities, are essential to this endeavor (e.g., \citealt{Hong2018bbhmergers, Hong+2020, Kremer2020catalog, Wu+2025dragon3, Mestichelli2026imbhs}; S.\ Agrawal et al., in prep.). 
The \catalog\ \citep[][hereafter K20]{Kremer2020catalog} is one of the most extensive such grids published to date and has become a widely utilized resource on the topics of stellar remnants and cluster dynamics. 
A major strength of these simulations is that, for a set of fiducial initial conditions, 
they reproduce many characteristics of old GCs inhabiting the Galactic halo. 
However, given that the properties of GCs at birth remain highly uncertain, these models represent only one plausible version of GC evolution. 

Besides dwelling in cluster environments, virtually all BH progenitors are found in multi-star systems, mainly binaries or triples 
(see the review by \citealt{Offner2023binaries} and references therein). 
These sustain their own rich interplay of orbital dynamics and stellar evolution, 
of which BBH formation is one possible outcome among many \citep[see review by][]{Marchant2024review}. 
In an isolated binary system, a BBH merger occurs only after multiple stages of stellar interaction, none of which can be said to be exhaustively understood. 
In a triple system or a dense cluster, dynamical interactions vastly expand the space of possible outcomes. 
In view of these considerations, theoretical studies are increasingly moving beyond the view of massive binary evolution and cluster dynamics as mutually exclusive channels for BBH formation \citep[e.g.,][]{Mapelli+2022, ArcaSedda+2023, Paiella+2025}. 
As the number of observed BBH mergers increases, so too does the need to consider the potential intersection between these channels. 
The \catalog\ is quite conservative in this regard, setting the initial binary fraction to be $5\%$ for all stars in GCs \citepalias{Kremer2020catalog}. 
Although this figure is typical of GCs as we now observe them \citep[e.g.,][]{Milone2012gcbinaryfraction, Ji2015_gcbinaries}, 
such a low {\it initial} binary fraction would be at odds with the demographics of young stellar populations in virtually all well-studied environments -- especially for massive stars \citep[e.g.,][]{Sana+2012, Sana2013_30Dor, Sana2025bloem, MoeDiStefano2017, Gautam2024galcentbinaries}.

In this paper, we present a new set of cluster models based on a new suite of {\tt Cluster Monte Carlo} (\cmc) simulations. 
Our models feature updated binary stellar evolution prescriptions and a realistic initial binary population based on observationally well-studied stellar populations in the local universe. 
In \S\ref{s:CMC}, we summarize our methods and characterize the channels by which BBHs form in our simulations. 
In \S\ref{s:Results}, we calculate the rate and properties of BBH mergers derived from a population of dense star clusters throughout the cosmos for a range of plausible cluster-formation histories; 
we also compare our results with previously published predictions based on the \catalog, 
highlighting the most important differences that arise from the presence of a larger population of primordial binary star systems. 
In \S\ref{s:Discussion}, we compare our results to those of previous theoretical and observational studies 
and discuss avenues for further improvement upon our models. 
We summarize our major findings in \S\ref{s:Summary}. 

Where required, we adopt a flat $\Lambda$CDM cosmology with local expansion rate $H_{0} = 67.7 \, \kms \,{\rm Mpc}^{-1}$ and density parameters $\Omega_{\rm m} = 1 - \Omega_{\rm \Lambda} = 0.31$ \citep{PlanckCollab2016}. 

\section{Star cluster models} \label{s:CMC}

\subsection{Methods} \label{s:CMC:methods}

We have conducted $24$ full-lifetime $N$-body simulations of dense star clusters with \cmc, a H\'{e}non-type Monte Carlo code for collisional stellar dynamics in spherical systems 
\citep{Henon1971montecarlo, Henon1971montecarlo2, Joshi2000, Joshi2001, Pattabiraman2013, Rodriguez2015, Rodriguez2022cmcreview}. 
\cmc\ includes prescriptions for all processes relevant for BBH formation in dense cluster environments, 
such as star-by-star single and binary evolution via \cosmic\ \citep{Breivik2020cosmic}, 
direct integration of strong binary--single and binary--binary encounters with {\tt FEWBODY} (\citealt{Fregeau2007}; plus 2.5PN terms for BHs as described in \citealt{Rodriguez2018cmc2.5pn, Rodriguez2018cmc2.5pn2}), 
binary formation via gravitational-wave capture and unbound three-body encounters, 
and global two-body relaxation in the presence of an external tidal field. 
Much of our input physics is the same as in the models of the \catalog\ \citepalias[][]{Kremer2020catalog}. 
Here we describe the few most important changes implemented since that work and their effects on our results.

The version of \cosmic\ currently linked to \cmc\ includes numerous refinements and expansions of the routines used in \citetalias{Kremer2020catalog}.
The main changes affecting BHs in our simulations are as follows: 
\begin{enumerate}
    \item We use the delayed-explosion recipe of \citet{Fryer2012remnants} to determine remnant masses for core-collapse supernovae (SNe), rather than the rapid-explosion recipe. 
    The delayed recipe notably predicts a continuous remnant mass spectrum for $\sim 3 \mbox{--} 5 \MSol$ (as opposed to one with a gap), 
    as favored by LVK observations \citep{Abac+2024} 
    and Galactic microlensing events \citep{Wyrzykowski2016}.

    \item We calculate the remnant masses of (pulsational) pair-instability SNe ([P]PISNe) following \citet{Marchant2019ppisn}. 
    This predicts a gap in the BH mass spectrum from isolated single-star evolution between $44$ and $120 \MSol$. 
    Accordingly, we refer to this mass domain as the ``PISN gap,'' 
    bearing in mind that BHs of these masses (or larger) may form in our models from the collapse of stellar collision products \citep[e.g.,][]{Kremer2020massgap, GonzalezPrieto+2021, GonzalezPrieto+2024}. 
    The \catalog\ utilized the prescription of \citet{Belczynski+2016}, 
    placing the lower edge of the PISN gap at $40.5 \MSol$.

    \item We determine the stability of mass transfer in binaries undergoing Roche lobe overflow following the {\tt COMPAS} code \citep{Neijssel2019stablemt}, 
    whereas the \catalog\ used criteria from \citet{Claeys2014_stablemt}. 
    Notably, the {\tt COMPAS} prescription assumes that mass transfer with stripped He star donors is dynamically stable. 
    This affects the properties of in-cluster BBH mergers that occur without dynamical intervention; 
    in particular, it reduces the proportion of mergers with very short delay times, which preferentially form via CE evolution \citep[see][]{GallegosGarcia2021bbh}.
\end{enumerate}

Our model grid explores the clusters' initial total number of bound stars $N/10^{5} = \{ 4, 8, 16 \}$, initial virial radius $r_{\rm v} / {\rm pc} = \{ 0.5, 1, 2 \}$, and metallicity $Z/Z_{\odot} = \{ 0.03, 0.3, 1\}$ with $Z_{\odot} = 0.0142$. 
We consider all combinations over these values except those with both $N/10^{5} = 16$ and $r_{\rm v} = 0.5 \pc$, 
which result in collisional runaways and thus cannot be evolved for a Hubble time using \cmc\ \citep{Kremer2020catalog, GonzalezPrieto+2021, GonzalezPrieto+2024}. 
All stellar masses are sampled from the canonical Kroupa initial mass function from $0.08$ to $150 \MSol$ \citep{Kroupa2001imf}, 
yielding an average stellar mass $\bar{m} \simeq 0.6 \MSol$ and hence initial cluster masses of $M_{0} / (10^{5} \MSol) = \{ 2.4, 4.8, 9.6 \}$. 
The initial stellar density profile follows a King model with dimensionless central potential $w_{0} = 5$ \citep{King1966}.

The initial binary population is modeled after the 
binary populations observed in the Solar neighborhood and nearby star-forming regions, covering the full main sequence up to $50 \MSol$ (see \citealt{Offner2023binaries} and references therein). 
We assign secondary stars to primordial binary systems from the predetermined set of cluster members, one at a time and without replacement, 
in such a way as to closely approximate the desired mass-ratio distribution and mass-dependent binary fraction 
without modifying the overall initial mass function \citep[see][]{KhuranaChatterjee2025}.
The initial binary fraction is given as a function of primary stellar mass as follows: 
For $m_{1} < 0.55 \MSol$ and $m_{1} > 50 \MSol$, we set $F_{\rm b} = 0.15$ and $F_{\rm b} = 0.82$, respectively. 
For masses $0.55 \MSol \leq m_{1} \leq 50 \MSol$, we interpolate continuously between these extremes with a logarithmic ramp:
\begin{equation}
    F_{\rm b}(m_{1}) = 0.15 + 0.34 \log_{10}(m_{1}/0.55 \MSol).
\end{equation}
This prescription mimics the observed close binary fraction data compiled by \citet{Offner2023binaries}, 
defined as the fraction of stars of a given mass with a companion closer than $10 \AU$.
For binaries with primary stars below $15 \MSol$, we sample initial semi-major axes from the log-normal separation distribution of \citet{Raghavan+2010} up to a cutoff at $10 \AU$. 
We draw initial orbital eccentricities from a uniform distribution, in accordance with {\it Gaia}'s astrometric characterization of close binaries \citep{Hwang+2022gaiaeccentricities}. 
Meanwhile, we draw the initial periods and eccentricities of binaries with primaries above $15 \MSol$ from the distributions reported for O-type stars by \citet{Sana+2012}. 
In addition to the mass-dependent binary fraction, a qualitatively important improvement is the initial presence of dynamically ``soft'' binaries, particularly at low stellar masses. 
Although soft binaries tend to dissolve over the lifetime of the cluster \citep{Heggie1975}, their presence at early times is important 
because they absorb a significant portion of the energy released by ``BH burning'' \citep{Kremer2020bhburning}; 
in this way, they stimulate the dynamical formation of hard BBHs (see \citealt{Wang+2022} and \citealt{OConnor+2026}, submitted). 

Another major methodological change affecting BH dynamics pertains to three-body binary formation for encounters between three initially unbound BHs.
This is a crucial component of BH burning and regulates the long-term dynamical evolution of GCs \citep[e.g.,][]{BreenHeggie2013, Morscher2013bhdynamics, Morscher2015bhdynamics, Kremer2019corecollapse, Kremer2020bhburning}. 
Our prescription for three-body binary formation is based on that of \citet{Morscher2013bhdynamics, Morscher2015bhdynamics}, 
with modifications motivated by more recent analytical and numerical studies of the process by \citet{GinatPerets2024} and \citet{Atallah+2024}. 
We define the dimensionless binding energy of a binary relative to the average kinetic energy of nearby objects, frequently called the ``hardness,'' as
\begin{equation}
    \eta = \frac{2 |E_{\rm b}|}{\langle m v^{2} \rangle},
\end{equation}
where $m$ and $v$ are the mass and velocity of background objects and $\langle \cdot \rangle$ indicates a suitable defined local average.
The rate of ``hard'' ($\eta \geq 1$) binary formation per interval of $\eta$ is given schematically by:
\begin{equation}
    \frac{\dif \Gamma}{\dif \eta} \sim \frac{G^{5} (m_{1}+m_{2})^{5} n^{3}}{\sigma^{9} \eta^{\gamma}} (1 + 2 \eta) \left( 1 + \frac{2 \eta (m_{1}+m_{2})}{m_{1}+m_{2}+m_{3}} \right),
\end{equation}
where $n$ is the local number density, $m_{1,2,3}$ are the masses involved in a three-body encounter that forms a binary containing $m_{1}$ and $m_{2}$, 
and the bracketed factors account for gravitational focusing. 
We set the exponent $\gamma = 3$ 
(\citealt{GoodmanHut1993}, \citealt{Atallah+2024}; cf.\ $\gamma = 5.5$ in \citealt{Morscher2013bhdynamics, Morscher2015bhdynamics}) 
and select a numerical coefficient to match the numerically determined total rate of steady-state hard binary formation from the aforementioned studies. 
We allow new binaries to have $\eta \in [1, 50)$, as in \citetalias{Kremer2020catalog}. 
Additionally, whereas the prescription implemented by \citet{Morscher2013bhdynamics, Morscher2015bhdynamics} 
always chooses the two most massive bodies in an encounter to form the new binary, 
we select the new binary's components at random; 
this change is motivated by the results of \citet{Atallah+2024} on pairing probabilities in unequal-mass encounters. 
This last modification has a small effect in practice, since newly formed BBHs are efficiently reprocessed by subsequent scattering events; 
nonetheless, it is a crucial correction because it allows a fuller exploration of possible mass ratios in dynamically assembled BBH mergers. 

\subsection{Auxiliary isolated binary dataset} \label{s:CMC:Methods:COSMICauxiliary}

In addition to our detailed cluster simulations, we used \cosmic\ to simulate the evolution of massive binary systems in isolation, using identical settings to our cluster models. 
We simulated a separate set of binaries for each metallicity value in our grid, with $\sim 10^{6}$ binaries in each set. 
All binary initial conditions were sampled from the same distributions as our in-cluster binaries, except that we include only systems with primary masses above $15 \MSol$. 
Each set yields $\sim 10^{3}$ BBH mergers with delay times less than $13.8 \Gyr$ relative to star formation. 

This auxiliary dataset serves a dual purpose. 
First, it allows us to sample binary properties that are not saved in \cmc-\cosmic\ models at present, most importantly the change in the orientation of a binary's orbit due to a BH natal kick. 
Second, it enables an apples-to-apples comparison of the predicted rates and properties of field- and cluster-derived BBH mergers with our chosen input physics. 
Although predictions from more detailed binary-evolution model grids are available \citep[e.g.,][]{Bavera+2021, GallegosGarcia2021bbh, Fragos+2023, Andrews+2025}, 
these have not been incorporated into our cluster models at present.

\begin{deluxetable*}{c|ccc||c|c|cc|ccccc|cc}
\tabletypesize{\scriptsize}
\tablecaption{Overview of \cmc\ Model Grid and Selected Results on BBH Mergers. \label{tab:bbhraw_v2}}
\tablehead{I & II & III & IV & V & VI & VII & VIII & IX & X & XI & XII & XIII & XIV & XV \\ 
    & & & & & & \multicolumn{2}{c}{Prim. Bin. Evo.} & \multicolumn{2}{c}{Dynamical} & \multicolumn{3}{c}{GW Captures} & \multicolumn{2}{c}{High-Mass} \\ 
    Model No. & $N/10^{5}$ & $Z/Z_\odot$ & $r_{\rm v}/{\rm pc}$ & $\max(N_{\rm BH})$ & $N_{\rm GW}$ & Quasi-Iso. & Perturbed & 1G & 2G+ & 2-body & 3-body & 4-body & PISN & IMBH 
}
\startdata
1 & 4 & 0.03 & 0.5 & 366 & 65 & 1 & 0 & 59 & 5 & 0 & 1 & 5 & 7 & 1 \\
2 & 4 & 0.03 & 1 & 420 & 91 & 27 & 9 & 52 & 3 & 0 & 1 & 3 & 5 & 0 \\
3 & 4 & 0.03 & 2 & 355 & 75 & 24 & 15 & 35 & 1 & 0 & 1 & 2 & 1 & 0 \\
4 & 4 & 0.3 & 0.5 & 329 & 56 & 0 & 0 & 53 & 3 & 0 & 1 & 1 & 5 & 1 \\
5 & 4 & 0.3 & 1 & 375 & 76 & 2 & 13 & 60 & 1 & 0 & 3 & 4 & 2 & 0 \\
6 & 4 & 0.3 & 2 & 314 & 70 & 14 & 13 & 41 & 2 & 0 & 0 & 3 & 0 & 0 \\
7 & 4 & 1.0 & 0.5 & 187 & 38 & 0 & 0 & 34 & 4 & 0 & 1 & 2 & 2 & 0 \\
8 & 4 & 1.0 & 1 & 103 & 32 & 9 & 1 & 21 & 1 & 0 & 0 & 2 & 1 & 0 \\
9 & 4 & 1.0 & 2 & 105 & 33 & 17 & 0 & 15 & 1 & 0 & 0 & 0 & 0 & 0 \\ \hline
10 & 8 & 0.03 & 0.5 & 859 & 161 & 2 & 0 & 136 & 23 & 1 & 7 & 15 & 19 & 4 \\
11 & 8 & 0.03 & 1 & 858 & 193 & 66 & 0 & 112 & 15 & 0 & 4 & 7 & 13 & 0 \\
12 & 8 & 0.03 & 2 & 834 & 174 & 46 & 22 & 101 & 5 & 0 & 1 & 1 & 5 & 0 \\
13 & 8 & 0.3 & 0.5 & 732 & 149 & 0 & 0 & 132 & 17 & 0 & 6 & 18 & 13 & 0 \\
14 & 8 & 0.3 & 1 & 805 & 162 & 18 & 5 & 124 & 15 & 0 & 0 & 9 & 9 & 0 \\
15 & 8 & 0.3 & 2 & 702 & 151 & 45 & 10 & 86 & 10 & 0 & 2 & 0 & 2 & 0 \\
16 & 8 & 1.0 & 0.5 & 473 & 105 & 2 & 0 & 91 & 12 & 0 & 3 & 8 & 5 & 1 \\
17 & 8 & 1.0 & 1 & 368 & 86 & 12 & 1 & 63 & 10 & 0 & 0 & 3 & 0 & 0 \\
18 & 8 & 1.0 & 2 & 266 & 74 & 38 & 0 & 31 & 5 & 0 & 0 & 6 & 0 & 0 \\ \hline
19 & 16 & 0.03 & 1 & 1997 & 349 & 29 & 0 & 264 & 56 & 3 & 21 & 16 & 47 & 1 \\
20 & 16 & 0.03 & 2 & 1895 & 387 & 195 & 0 & 161 & 31 & 3 & 8 & 4 & 30 & 0 \\
21 & 16 & 0.3 & 1 & 1951 & 354 & 21 & 0 & 280 & 53 & 3 & 15 & 16 & 22 & 2 \\
22 & 16 & 0.3 & 2 & 1691 & 327 & 71 & 28 & 198 & 30 & 0 & 9 & 9 & 10 & 0 \\
23 & 16 & 1.0 & 1 & 1063 & 259 & 28 & 2 & 189 & 40 & 1 & 12 & 11 & 2 & 0 \\
24 & 16 & 1.0 & 2 & 724 & 160 & 57 & 1 & 84 & 18 & 0 & 0 & 2 & 0 & 0 \\
\enddata
\tablecomments{Column V refers to the maximum number of BHs retained in the cluster at any time. 
Column VI refers to the number of BBH mergers that occur within $13.8 \Gyr$ of cluster formation. 
Columns VII and VIII refer to mergers derived from a primordial binary pair that have experienced {\it zero} and {\it one or more} strong encounters with another BH prior to merger, respectively. 
Columns XI through XIII show the numbers of dynamically assembled mergers that occur through GW capture in two-, three-, and four-body scattering events. 
Column XIV refers to mergers with at least one component between $44 \MSol$ and $120 \MSol$ in mass, thus occupying the pair instability mass gap. 
Column XV refers to mergers with at least one component above $120 \MSol$.}
\end{deluxetable*}

\subsection{BBH merger yields and channels} \label{s:CMC:channels}

In all, we obtain 3,627 BBH mergers across our 24 cluster models within $13.8 \Gyr$ of cluster formation. 
Merger data relevant to the results presented in this work may be downloaded from an online repository.\footnote{\doi{10.5281/zenodo.20651303}}

Table \ref{tab:bbhraw_v2} lists the following quantities for each model: 
model index (column I), the global parameters $N$, $Z$, and $r_{\rm v}$ (columns II--IV);
$\max(N_{\rm BH})$, the maximum number of BHs retained in each cluster over its lifetime (V); 
and $N_{\rm GW}$, the total number of BBHs that merge within $13.8 \Gyr$ of cluster formation (VI). 
Columns VII--XV show a breakdown of the mergers produced by each model 
according to their formation channels and other properties, as described below. 

One of the clearest trends in the tabulated data is that the number of retained BHs per cluster 
decreases drastically with increasing $Z$, all else being equal. 
This is a well known effect, understood mainly as a result of stronger wind-driven mass loss from higher-metallicity stars \citep[e.g.,][]{Chatterjee2017, YeFishbach2024, Ye+2026}. 
However, when we compare our simulation outputs with similar models in the \catalog, 
we find that ours often retain fewer BHs by up to a factor of $\simeq 2$, especially at high metallicity. 
We have determined that the SN explosion recipe is the likely cause: 
By using the delayed SN recipe over the rapid, our models form a greater share of low-mass ($2.5 \mbox{--} 8 \MSol$) BHs. 
All else being equal, lower-mass BHs receive larger natal kicks in our models 
because we reduce kick velocities by a factor proportional to the SN fallback mass fraction (see \citetalias{Kremer2020catalog}); 
hence, the overall probability of retaining BHs in our simulations is moderately smaller. 
However, this does not lead to a lower BH merger rate 
because the \emph{retained} BHs in our simulations are \emph{larger} on average than those in the \catalog\ (see \S\ref{s:Results:m1_q}); 
thus, our models form merging BBHs more efficiently through dynamics, 
yielding an equal or greater cosmic merger rate to that derived from the \catalog\ (\S\ref{s:Results:RatesChannelsMasses}). 

We classify BBH mergers according to their formation channels in our models. 
\emph{Primordial binary evolution} comprises two sub-channels, 
which we term \emph{quasi-isolated} and \emph{perturbed}. 
The former consists of systems where a BBH forms via evolutionary processes internal to the binary system and merges without experiencing a strong encounter (i.e., a {\tt FEWBODY} call) in the cluster, 
i.e. purely through GW emission, much as in a putative field population. 
In the latter group, the BBH experiences at least one strong encounter between formation and merger, but the original pair survives with modified orbital properties. 
The number of mergers produced by these sister channels (Table \ref{tab:bbhraw_v2}, columns VII and VIII) is a sensitive function of the cluster's global parameters (see below). 

Previous studies have established and characterized 
many modes of \emph{dynamical formation} within clusters. 
The dominant mode is standard BH burning \citep{Kremer2020bhburning}, in which dynamically assembled hard BBHs 
shrink gradually through many successive scattering events, 
eventually merging via GW emission either within the cluster or after being dynamically ejected \citep{Rodriguez2015, Rodriguez+2016, Rodriguez2018cmc2.5pn, Rodriguez2018cmc2.5pn2}. 
Direct GW capture, where two BHs become tightly bound and initiate a rapid inspiral by emitting a burst of gravitational radiation during a relativistic encounter, 
can occur during resonant binary--single \citep{Samsing+2014} or binary--binary interactions \citep{Zevin+2019} 
as well as, more rarely, through close two-body encounters \citep{Samsing+2020}. 
We track each of these dynamical formation channels in our analysis and list the respective numbers of GW capture events leading to mergers in Table \ref{tab:bbhraw_v2}. 
However, for the most part, it suffices to split the dynamically formed mergers coarsely 
into \emph{first-generation (1G)} and {\it hierarchical mergers}, 
with the latter involving BHs that formed from previous in-cluster mergers (denoted 2G, 3G, and so on); 
these are also tallied in Table \ref{tab:bbhraw_v2}. 
We automatically determine the generational number of both BHs in each merger and analyze the rates and properties of 1G dynamical and hierarchical mergers separately. 
The hierarchical mergers in our models consist of 334 2G+1G pairings, 25 2G+2G pairings, and a single 3G+1G pairing.
Higher-generation BHs frequently receive a large enough GW recoil kick to escape a typical GC upon formation \citep{Rodriguez+2019}. 
For a recent study of higher-generation BH mergers in a \cmc\ simulation with $N = 10^{7}$, see \citet{Mai+2026}.

Finally, Table \ref{tab:bbhraw_v2} also lists the number of mergers involving a BH 
that lies in the nominal PISN mass gap ($44 \mbox{--} 120 \MSol$, column XI) 
or the intermediate-mass BH (IMBH) domain ($> 120 \MSol$, XII). 
As mentioned above, these BHs derive from over-massive stars formed via in-cluster stellar collisions \citep{Kremer2020massgap, GonzalezPrieto+2024}. 
All such mergers are thus classified as dynamically assembled. 
The largest BH to participate in a merger in our models had a mass of $391 \MSol$. 
Model number 10 formed the greatest number of mergers with IMBHs, at 4; unsurprisingly, this is also our densest and lowest-metallicity model. 
About half of all hierarchical mergers in our models fall in the PISN gap. 
However, most mergers involving IMBHs in our models are 1G+1G, due to a confluence of factors disfavoring long-term retention of $100 \mbox{--} 1000 \MSol$ BHs within GCs \citep{GonzalezPrieto+2022, Martinez+2026imbhsurvival}. 

We have obtained analytical formulae to estimate the typical number of mergers produced via each channel in a cluster of a given mass and radius at a given metallicity. 
For each channel, we consider two simple functional forms used in previous studies \citep[e.g.,][]{Hong2018bbhmergers, AntoniniGieles2020bbhpopsynth, Mai+2026}: 
A dual power-law in mass and radius
\begin{equation} \label{eq:NGW_power}
    N_{\rm GW} = A M_{6}^{\alpha} x^{\beta}, 
\end{equation}
where $M_{6} = M_{0} / (10^{6} \MSol)$ and $x = r_{\rm v} / (1 \pc)$; 
and a combination of a power law in mass and an exponential in radius:
\begin{equation} \label{eq:NGW_exp}
    N_{\rm GW} = A M_{6}^{\alpha} \exp(-x/l).
\end{equation} 
For primordial binary mergers (taking quasi-isolated and perturbed systems together), we find that the dual power-law yields a better overall fit 
in terms of lower r.m.s.\ residuals between the best fit and the \cmc\ data; 
for hierarchical mergers, equation \eqref{eq:NGW_exp} provides a much better fit \citep[consistent with][]{Mai+2026}. 
For 1G dynamical mergers, the best fits for each form are comparable in quality; 
we proceed with the best fit of equation \eqref{eq:NGW_exp}, which yields marginally lower r.m.s.\ residuals. 
Table \ref{tab:fittingformulae} lists the best-fitting parameters of each preferred model;
it also shows formal uncertainties computed under the assumption that $N_{\rm GW}$ for each \cmc\ model obeys Poisson statistics.

\begin{deluxetable}{c|c|cccc|c}
\tablecaption{Derived quantities related to BBH formation within GCs. \label{tab:fittingformulae}}
\tablehead{
    \colhead{$Z/Z_\odot$} & \colhead{Channel} & \colhead{$A$} & \colhead{$\alpha$} & \colhead{$\beta$} & \colhead{$l$} & \colhead{$K_{\rm GW}$}}
\startdata
    $0.03$ & Bin. & $3.6 \pm 0.7$ & $2.0 \pm 0.3$ & $2.2 \pm 0.2$ & -- & $2.5$ \\ 
    & 1G Dyn. & $361 \pm 18$ & $1.2 \pm 0.1$ & -- & $3.1 \pm 1.6$ & $5.8$ \\
    & Hier. & $140 \pm 29$ & $2.1 \pm 0.1$ & -- & $1.3 \pm 0.2$ & $2.3$ \\ \hline
    $0.3$ & Bin. & $1.3 \pm 0.5$ & $1.9 \pm 0.4$ & $2.7 \pm 0.3$ & -- & $2.6$ \\ 
    & 1G Dyn. & $370 \pm 18$ & $1.2 \pm 0.1$ & -- & $3.6 \pm 2.9$ & $5.8$ \\
    & Hier. & $87 \pm 15$ & $1.9 \pm 0.2$ & -- & $2.2 \pm 1.5$ & $2.6$ \\ \hline
    $1.0$ & Bin. & $1.3 \pm 0.5$ & $2.4 \pm 0.4$ & $2.1 \pm 0.3$ & -- & $2.0$ \\ 
    & 1G Dyn. & $371 \pm 24$ & $1.5 \pm 0.1$ & -- & $1.4 \pm 0.1$ & $3.8$ \\
    & Hier. & $81 \pm 21$ & $2.0 \pm 0.2$ & -- & $1.4 \pm 0.4$ & $2.5$
\enddata
\end{deluxetable}





\begin{figure*}
    \centering
    \includegraphics[width=\textwidth]{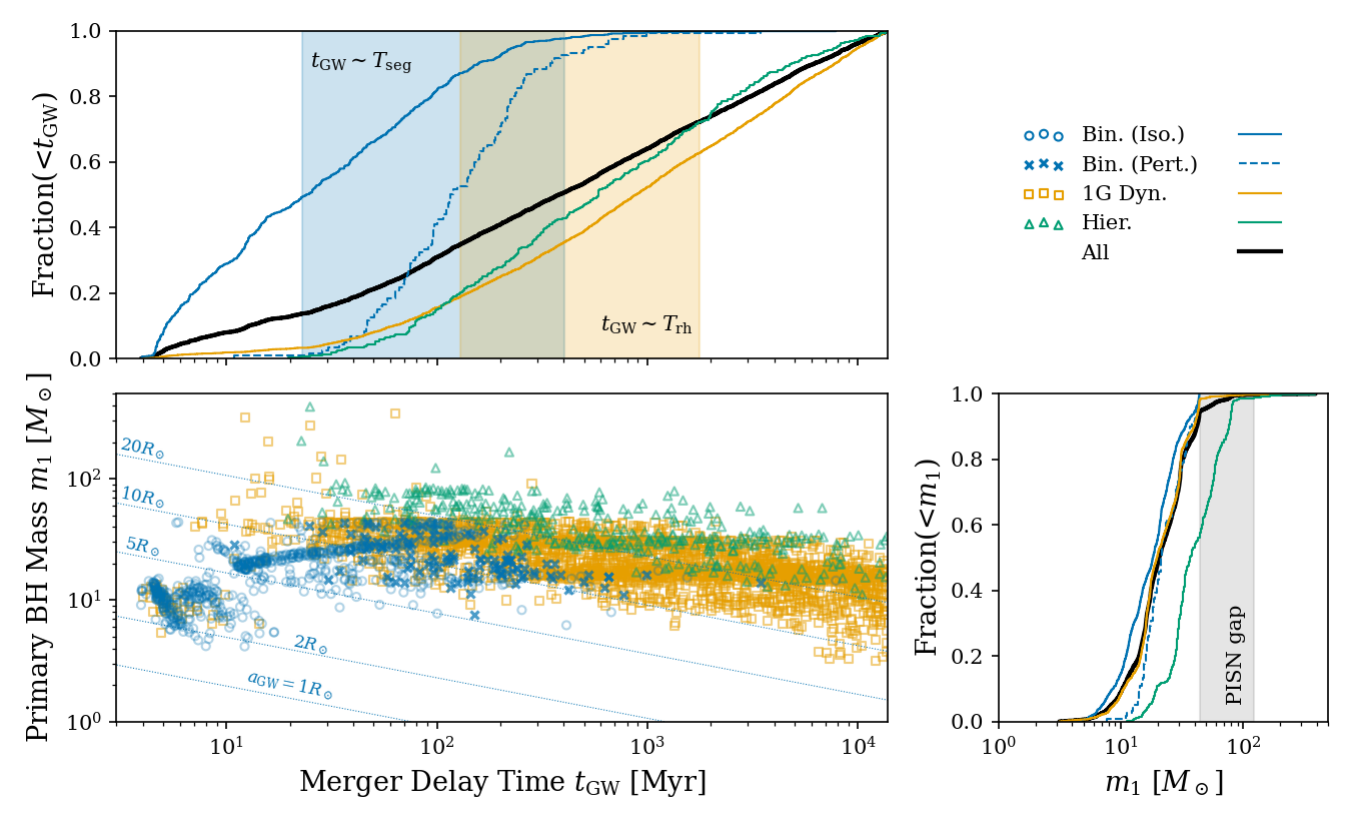}
    \caption{{\it Scatter plot:} Distribution of primary BH mass $m_{1}$ versus delay time $t_{\rm GW}$ for all BBH mergers in our \cmc\ models. 
    Different merger channels are indicated by marker shape and color: blue circles and crosses for isolated and perturbed primordial binary evolution, respectively; 
    amber squares for 1G dynamical assembly; and green triangles for hierarchical mergers. 
    The dotted blue lines mark curves of constant semi-major axis on which a circular, equal-mass binary has inspiral time equal to $t_{\rm GW}$.
    {\it Upper and right panels:} Marginal cumulative distributions of $t_{\rm GW}$ and $m_{1}$, respectively: solid blue and dashed blue for isolated and perturbed binary evolution, amber for 1G dynamics, and green for hierarchical mergers. 
    The blue and amber shaded regions in the upper panel indicate respectively the ranges where $t_{\rm GW}$ is comparable to the mass segregation time $T_{\rm seg}$ for stellar-mass BHs and the host cluster's relaxation time $T_{\rm rlx}$. 
    The grey region on the right indicates the nominal PISN mass gap.} 
    \label{fig:masses_and_delays}
\end{figure*}

A key finding of this work is that different modes of BBH formation operating within a single population of dense clusters produce mergers with significantly different properties and different correlations among them. 
Figure \ref{fig:masses_and_delays} illustrates this by showing the relation between the primary BH mass $m_{1}$ and the merger delay time $t_{\rm GW}$, 
the latter being the elapsed time since cluster formation at the moment of each merger. 
The main panel of the figure shows both quantities on a scatter plot, 
conveying which channels contribute mergers in a given interval of mass or parent cluster age; 
the two secondary panels show each quantity's marginal cumulative distribution function. 

Examining the delay-time distributions first, a readily apparent feature is the separation of timescales between primordially paired and dynamically assembled mergers, 
with the former having significantly shorter delay times. 
To understand this, it is helpful to introduce two characteristic timescales relevant to BH dynamics within clusters: 
the relaxation time
\begin{align} \label{eq:Trlx_def}
    T_{\rm rlx} &= \frac{0.1 N}{\ln{N}} \left( \frac{r_{\rm v}^{3}}{G M} \right)^{1/2} \sim 100 \Myr \mbox{--} 2 \Gyr, 
\end{align} 
which is a measure of the rate of the cluster's overall dynamical evolution; 
and the mass-segregation time,
\begin{equation} \label{eq:Tseg_def}
    T_{\rm seg} = \frac{m_{*}}{m_{\rm BH}} T_{\rm rlx} = 0.1 T_{\rm rlx} \left( \frac{m_{*}}{\MSol} \frac{10 \MSol}{m_{\rm BH}} \right), 
\end{equation}
which measures the time over which BHs (or massive stars) sink to the cluster's center due to dynamical friction. 
In the upper panel of Fig.\ \ref{fig:masses_and_delays}, we show the span of these timescales across our model grid as amber and blue shaded regions, respectively.
Primordial binary mergers mostly have $t_{\rm GW} \lesssim T_{\rm seg}$, and dynamical mergers (both 1G and hierarchical) tend to have $t_{\rm GW} \gtrsim T_{\rm seg}$, as expected. 
Moreover, within the category of primordial binary mergers, quasi-isolated systems merge faster than perturbed ones, with respective medians delay times of $\sim 25 \Myr$ and $\sim 100 \Myr$. 
These trends reflect a dynamical selection effect within dense cluster environments: 
only the tightest BBHs produced from massive binaries can merge in quasi-isolation, since the first strong encounter typically takes place after a time $\sim T_{\rm seg}$. 
Similarly, the perturbed-binary channel can operate only when $T_{\rm seg} \lesssim t_{\rm GW} \lesssim T_{\rm rlx}$: 
these systems have experienced at least one strong encounter, but not so many that exchanges or ionizations have become statistically inevitable. 

Another noticeable feature of the primordially paired mergers is that they show significant structure in the $t_{\rm GW}$--$m_{1}$ plane. 
Further insight on this aspect can be gained by calculating a characteristic orbital separation $a_{\rm GW}$, 
defined as the initial semi-major axis of a circular binary with component masses $m_{1} = m_{2} = m_{\rm BH}$ that merges purely through GW emission in a time $t_{\rm GW}$ \citep{Peters1964}: 
\begin{align} \label{eq:aGW_definition}
    a_{\rm GW} &\equiv \left( \frac{512}{5} \frac{G^{3} m_{\rm BH}^{3}}{c^{5}} t_{\rm GW} \right)^{1/4} \\ 
    &\simeq 6.0 \RSol \left( \frac{m_{\rm BH}}{10 \MSol} \right)^{3/4} \left( \frac{t_{\rm GW}}{100 \Myr} \right)^{1/4}. \nonumber
\end{align}
This quantity gives a rough estimate of the orbital separation just after the formation of the secondary BH, accurate within a factor of $\simeq 2$ in most cases. 
In the main panel of Fig.\ \ref{fig:masses_and_delays}, we plot several contours of constant $a_{\rm GW}$ as thin, dotted blue lines. 
We see that the blue data have a roughly bimodal distribution with respect to $a_{\rm GW}$, with one group concentrated at $2 \RSol \lesssim a_{\rm GW} \lesssim 4 \RSol$, another at $5 \RSol \lesssim a_{\rm GW} \lesssim 20 \RSol$, and a distinct gap between them; 
within these groups, there are hints of further substructure. 
This distribution presumably reflect different modes of binary interaction leading to BBH formation, 
with phases of common-envelope evolution and stable mass transfer occurring in various combinations \citep[see, e.g.,][]{Broekgaarden+2026}. 
However, this aspect of the results should not be taken too literally, 
as rapid population synthesis codes such as \cosmic\ are known to predict significantly different distributions of BBH properties 
from grids of detailed binary evolution models \citep[e.g.,][]{Bavera+2021, GallegosGarcia2021bbh}. 

As the properties of 1G dynamical and hierarchical mergers within dense clusters have been explored extensively in previous works, 
we remark only on a few key points where our results differ.  
Broadly speaking, less-massive BHs have longer $t_{\rm GW}$; 
this is because the few largest BHs present in the cluster at any time dominate the BH burning process, 
forcing smaller BHs to ``wait their turn'' to pair off and merge \citep[see][]{YeFishbach2024}. 
This trend also exists in the \catalog\ models \citepalias{Kremer2020catalog}; it is more pronounced in ours because, 
all else being equal, having a higher binary fraction increases the rate of strong scattering events. 
Another novel result compared to \citetalias{Kremer2020catalog} is that hierarchical mergers formed in our models have moderately \emph{shorter} delay times than 1G dynamical mergers, 
with a median $t_{\rm GW} \sim 600 \Myr$ versus $\sim 1 \Gyr$. 
Among mergers in the \catalog, the delay time distributions of these two groups are indistinguishable. 
In our models, the main cause of this difference is that there are more 2G BHs present at early times in our models, these being derived in turn from primordial binary systems with short delay times. 
Interestingly, \citet{Mai+2026} have reported a similar offset between 1G and hierarchical mergers in a 10-million-body cluster model with a {\it low} binary fraction. 
We leave the task of developing a more detailed understanding of these trends for future work.

\section{Predicted merger rates and demographics} \label{s:Results}

\subsection{Cluster population synthesis} \label{s:CMC:population}

To predict the cosmological rate and demographics of BBH mergers from our simulation results, 
we weight each merger according to the rate of GC formation 
and the probability of forming a parent cluster of a given mass, radius, and metallicity as functions of redshift \citep[see, e.g.,][]{FishbachFragione2023,YeFishbach2024,Ye+2026}. 
We adopt the following fiducial assumptions:
\begin{itemize}
    \item \emph{Mass:} We assume that GC birth mass $M_{0}$ is distributed on the interval $[10^{4}, 10^{8}] \MSol$ according to a Schechter-type function \citep[e.g.,][]{ProtegiesZwart2010ymcreview,AntoniniGieles2020prd}:
    \begin{equation} \label{eq:birthmass_distribution}
        \phi_{M0}(M_{0}) \, \dif M_{0} \propto M_{0}^{-2} \exp(-M_{0}/M^{\star}) \, \dif M_{0},
    \end{equation}
    with Schechter mass $M^{\star} = 10^{6.26} \MSol$.

    \item \emph{Radius:} We assume that the birth virial radius $r_{\rm v}$ follows a log-normal distribution:
    \begin{equation} \label{eq:rv_distribution}
        \phi_{r}(r_{\rm v}) \, \dif r_{\rm v} \propto \exp\left[ - \frac{\log^{2}(r_{\rm v}/\mu_{r})}{2 \sigma_{r}^{2}} \right] \dif \log{r_{\rm v}}
    \end{equation}
    with median $\mu_{r} = 1 \pc$ and dispersion $\sigma_{r} = 0.3 \, {\rm dex}$. 
    This differs from the Gaussian radius distribution assumed by previous studies \citep{FishbachFragione2023, YeFishbach2024, Ye+2026}, 
    a change motivated by (i) the broader $r_{\rm v}$ distribution inferred from GWTC-3.0 data by \citet{FishbachFragione2023} 
    and (ii) the observed radius distribution of present-day GCs \citep{Harris1996} and YMCs \citep{ProtegiesZwart2010ymcreview}. 
    The difference amounts to a slight adjustment of the relative weights of models with different $r_{\rm v}$ 
    and has little impact on our main conclusions.

    \item \emph{Metallicity:} We assume that clusters formed at redshift $z$ have a log-normal metallicity distribution:
    \begin{equation} \label{eq:metallicity_distribution}
        \phi_{\zeta}(\zeta; z) \, \dif \zeta \propto \exp\left[ - \frac{(\zeta - \mu_{\zeta}(z))^{2}}{2 \sigma_{\zeta}^{2}} \right] \dif \zeta
    \end{equation}
    where $\zeta \equiv \log(Z/Z_{\odot})$.
    We assume the location of the peak evolves with redshift as \citep[e.g.,][]{MadauFragos2017} 
    \begin{equation} \label{eq:metallicity_evolution}
        \mu_{\zeta}(z) = 0.153 - 0.074 z^{1.34}
    \end{equation}
    and that the dispersion $\sigma_{\zeta} = 0.5 \, {\rm dex}$ is independent of redshift. 

    \item {\it Cluster birth rate:} We model the comoving GC formation rate density following \citet{MadauFragos2017}:
    \begin{equation} \label{eq:SFR}
        \mathcal{R}_{\rm GC}(z) = \frac{B \left[ (1+z)/(1+z_{\rm p}) \right]^{a_{z}}}{1+\left[ (1+z) / (1+z_{\rm p}) \right]^{a_{z} + b_{z}}},
    \end{equation} 
    where $z_{\rm p}$ is the redshift at which the function peaks and $a_{z}$ and $b_{z}$ are shape parameters. 
    We assume that this formula holds up to a maximum redshift $z_{\rm max} = 20$ and that $\mathcal{R}_{\rm GC}(z > z_{\rm max}) = 0$. 
    Unless otherwise specified, we set $z_{\rm p} = 2.2$, $a_{z} = 2.6$, and $b_{z} = 3.6$, so that the GC formation rate traces the overall cosmic star formation history. 
    The normalization factor $B$ is chosen to match the observed GC number density in the local universe \citep[e.g.,][]{Harris+2013gccounts, Rodriguez2015}:
    \begin{equation} \label{eq:nGC_integral}
        n_{\rm GC} = \int_{0}^{t_{\rm max}} \mathcal{R}_{\rm GC}(t) \, \dif t = 2.1 \, {\rm Mpc}^{-3},
    \end{equation}
    where $t_{\rm max}$ is the lookback time at redshift $z_{\rm max} = 20$.
\end{itemize}

Under these assumptions, the contribution of each BH merger in the model grid to the cosmic merger rate at redshift $z$ is
\begin{align}
    \mathcal{R}_{ijk}(z) &= \phi_{M0}(M_{0}^{(i)}) \Delta_{M,i} \, \phi_{r} (r_{\rm v}^{(j)}) \, \Delta_{r}^{(j)} \, K_{\rm GW}(Z^{(k)}) \nonumber \\
    & \times \sum_{l} \mathcal{R}_{\rm GC}(z_{t}(t_{z}+\tau_{l})) \, \phi_{\zeta}(\zeta(Z^{(k)}); z_{t}(t_{z}+\tau_{l})) \Delta_{Z}^{(k)}, \label{eq:weighting_equation}
\end{align} 
and the combined rate is $\mathcal{R}(z) = \sum_{i,j,k} \mathcal{R}_{ijk}(z)$, summing over all BBH mergers in the grid.
Here, $\Delta_{M}^{(i)}$, $\Delta_{r}^{(j)}$, and $\Delta_{Z}^{(k)}$ are the dimensions of the $(i,j,k)$-th grid cell. 
The quantity $K_{\rm GW}(Z)$ is a correction factor that accounts for the tidal evaporation of clusters prior to the present day (see below). 
Other quantities include $t_{z}$, the lookback time at redshift $z$; and $\tau_{l}$, the delay time of the $l$th merger formed in a cluster with these properties. 
The intermediate function $z_{t}(\cdot)$ is the inverse of the relation between $z$ and $t_{z}$. 
We carry out this calculation separately for primordial binary, 1G dynamical, and hierarchical mergers.  

The correction factor $K_{\rm GW}(Z)$ is defined as the ratio of the number of BBH mergers produced in \emph{all} clusters formed over the history of the universe to the number formed in clusters \emph{surviving today}. 
We calculate this quantity following \citet{AntoniniGieles2020prd}, who estimated the average mass (denoted $\Delta$) each cluster must lose to tidal evaporation (in addition to stellar evolution) 
in order to obtain the observed GC mass/luminosity function from a population of clusters whose birth mass function obeys equation \eqref{eq:birthmass_distribution}. 
Under their model, the present-day GC mass function is
\begin{equation}
    \phi_{M}(M; \Delta) \propto (M + \Delta)^{-2} \exp\left( - \frac{M + \Delta}{(M^{\star}/2)} \right)
\end{equation}
with $\Delta = 10^{5.33} \MSol$;
the underlying mapping between initial and present-day GC mass is $M_{0} = 2(M + \Delta)$, 
implying that only clusters born with $M_{0} \gtrsim 2 \Delta$ can survive for a Hubble time. 
If the average number of BBH mergers formed in a given cluster is $N_{\rm GW} = N_{\rm GW}(M_{0}; Z)$, then we have
\begin{equation}
    K_{\rm GW}(Z) = \frac{\int \phi_{M0}(M') N_{\rm GW}(M';Z) \, \dif M'}{\int \phi_{M}(M'/2) N_{\rm GW}(M';Z) \, \dif M'}.
\end{equation}
Both integrals are evaluated over $M' \in [10^{4}, 10^{8}] \MSol$. 
Since the form of $N_{\rm GW}$ varies by channel, 
we calculate a different $K_{\rm GW}$ for the primordial binary, 1G dynamics, and hierarchical channels at each metallicity; 
the resulting values are listed in Table \ref{tab:fittingformulae}. 

\subsection{Rates by channel and mass} \label{s:Results:RatesChannelsMasses}

\begin{figure*}
    \centering
    \includegraphics[width=\linewidth]{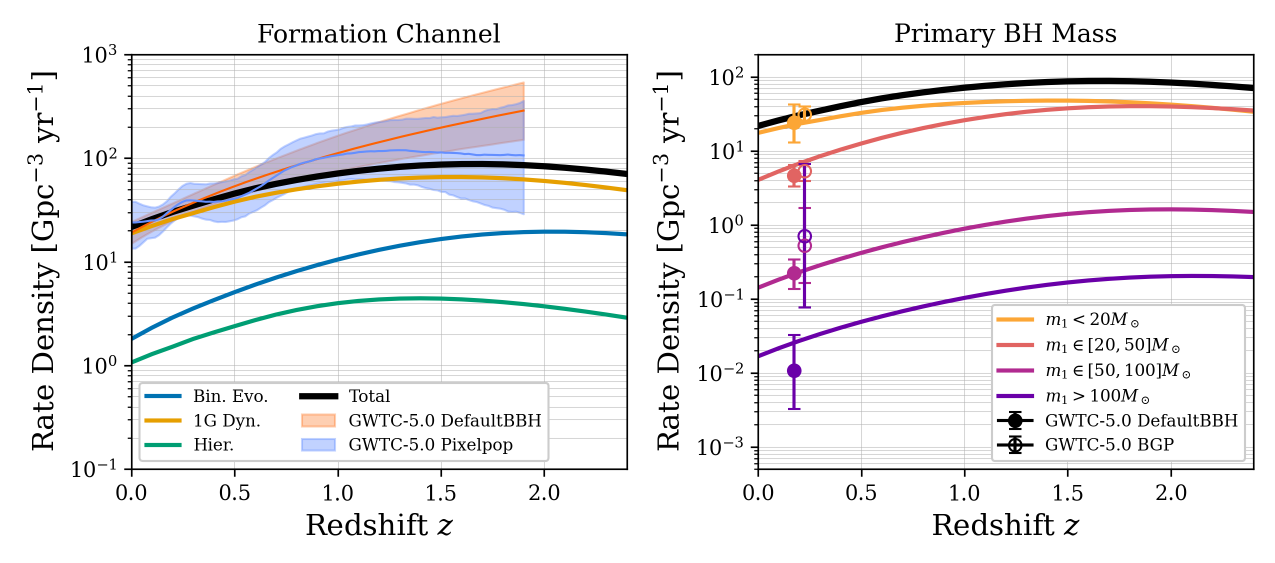}
    \caption{Predicted redshift evolution of the rate of cluster-derived BBH mergers under our fiducial cosmic GC formation history. 
    In both panels, the black curve indicates the predicted total cluster-derived merger rate. 
    Note that the vertical dynamic range differs between panels.
    {\it Left:} Thick colored curves indicate the merger rate broken down by formation channel: 
    primordial binary evolution (blue), 1G+1G dynamical assembly (amber), and hierarchical mergers (green). 
    The shaded regions indicate the central 90\% credible region inferred from LVK observations in GWTC-5.0 under two inference models: 
    the strongly modeled {\sc Default BBH}, which assumes power-law evolution as in equation \eqref{eq:R_powerlaw}; 
    and the weakly modeled {\sc Pixelpop} (light red).
    {\it Right:} The colored curves show the total merger rate binned by primary BH mass as follows, with lighter (darker) shades for smaller (greater) masses: $m_{1} < 20 \MSol$, $m_{1} \in [20,50] \MSol$, $m_{1} \in [50,100] \MSol$, and $m_{1} > 100 \MSol$. 
    The similarly colored points represent the median observational rates at $z = 0.2$ within the same bins, 
    derived by integrating the GWTC-5.0 posteriors for the {\sc Default BBH} (filled circles; \citealt{LVK2026_populations}) 
    and {\sc Binned Gaussian Process} inference models (empty circles; A. Ray, private communication) models.
    Vertical error bars indicate the central $90\%$ credible interval. 
    Note that the GWTC-5.0 data have been horizontally shifted from $z=0.2$ by a small amount for visual clarity.}
    \label{fig:rate_redshift}
\end{figure*}

Figure \ref{fig:rate_redshift} shows the predicted rate of cluster-derived BBH mergers under our fiducial assumptions as a function of redshift out to $z = 2$. 
In the left-hand panel, we show the total rate and the contributions of each formation channel. 
We also show the median and central 90\% credible region of the cosmic merger rate 
as inferred from the sample of 259 likely BBH mergers cataloged in GWTC-5.0 under the {\sc Default BBH} and {\sc PixelPop} population models \citep{LVK2026_populations}. 
In the right-hand panel, we plot the merger rate in the following bins of primary BH mass $m_{1}$: 
$m_{1} \leq 20 \MSol$, $20 \MSol < m_{1} \leq 50 \MSol$, $50 \MSol < m_{1} \leq 100 \MSol$, and $m_{1} > 100 \MSol$. 
We also plot observational estimates of the merger rates in the same bins at $z = 0.2$, 
computed by integrating the GWTC-5.0 {\sc Default BBH} and {\sc Binned Gaussian Process} posterior distributions over these bins.


\begin{deluxetable*}{l|cc|cccc}
\tabletypesize{\scriptsize}
\tablecaption{Summary of Cluster-Derived BBH Merger Rates and Demographics. \label{tab:rates}}
\tablehead{
 \colhead{Channel or Mass Bin} & \colhead{$\mathcal{R}_{0}/ ({\rm Gpc}^{-3} \yr^{-1})$} & \colhead{$\kappa$} & $m_{1}/\MSol$ & $q$ & $\chi_{\rm eff}$ & $\chi_{\rm p}$ 
}
\startdata
Bin. Evo. & $1.8$ & $2.5$ & $12.1^{+9.4}_{-5.7}$ & $0.74^{+0.22}_{-0.37}$ &  $0.11^{+0.10}_{-0.17}$ & $0.01^{+0.20}_{-0.01}$ \\ 
1G Dyn. & $19.5$ & $1.6$ & $15.3^{+15.4}_{-7.5}$ & $0.82^{+0.17}_{-0.38}$ & $0.00^{+0.13}_{-0.12}$ & $0.14^{+0.12}_{-0.11}$ \\ 
Hier. & $1.1$ & $1.9$ & $27.8^{+27.6}_{-17.7}$ & $0.54^{+0.41}_{-0.18}$ & $0.01^{+0.40}_{-0.42}$ & $0.58^{+0.10}_{-0.43}$ \\ \hline 
$m_{1} < 20 \MSol$ & $18.3$ & $1.3$ & -- & $0.83^{+0.15}_{-0.39}$ & $0.01^{+0.15}_{-0.13}$ & $0.14^{+0.12}_{-0.14}$ \\ 
$m_{1} \in [20,50] \MSol$ & $4.2$ & $2.7$ & -- & $0.73^{+0.25}_{-0.32}$ & $0.01^{+0.19}_{-0.17}$ & $0.14^{+0.51}_{-0.10}$ \\
$m_{1} \in [50,100] \MSol$ & $0.14$ & $2.7$ & -- & $0.48^{+0.20}_{-0.31}$ & $-0.03^{+0.38}_{-0.31}$ & $0.26^{+0.42}_{-0.23}$ \\ 
$m_{1} > 100 \MSol$ & $0.017$ & $2.6$ & -- & $0.28^{+0.67}_{-0.26}$ & $0.04^{+0.11}_{-0.04}$ & $0.18^{+0.46}_{-0.12}$ \\ 
\hline
{\bf All} & $\mathbf{22.4}$ & $\mathbf{1.7}$ & $\mathbf{15.3^{+15.6}_{-7.4}}$ & $\mathbf{0.81^{+0.18}_{-0.38}}$ & $\mathbf{0.01^{+0.15}_{-0.13}}$ & $\mathbf{0.14^{+0.14}_{-0.12}}$
\enddata
\tablecomments{The reported rate parameters $\mathcal{R}_{0}$ and $\kappa$ are obtained by fitting equation \eqref{eq:R_powerlaw} to the merger rates 
derived from our \cmc\ models at $z \leq 1$ under our fiducial cluster formation history. 
The quoted values and error bars for $m_{1}$, $q$, $\chi_{\rm eff}$, $\chi_{\rm pre}$ are weighted medians and central $90\%$ intervals evaluated at $z = 0.2$.}
\end{deluxetable*}


Our fiducial prediction of the total rate is in broad agreement with LVK observations, 
lying within or barely outside the 90\% credible interval for both distributions out to $z \simeq 1$. 
To make a quantitative comparison, we adopt the widespread power-law parametrization 
\begin{equation} \label{eq:R_powerlaw}
    \mathcal{R}(z) = \mathcal{R}_{0} (1 + z)^{\kappa}.
\end{equation} 
We have fitted this formula to the derived merger rates for each channel and mass bin in the domain $z < 1$, 
listing the best-fitting parameters in Table~\ref{tab:rates}. 
The table also gives weighted medians and central 90\% quantile intervals 
for the key observables $m_{1}$, $q$, $\chi_{\rm eff}$, and $\chi_{\rm p}$ 
for each formation channel and mass bin at $z = 0.2$ (see Sections \ref{s:Results:m1_q} and \ref{s:Results:spins}). 
We obtain a total merger rate density of $\mathcal{R}_{0} = 22.4 \, {\rm Gpc}^{-3} \, \yr^{-1}$ at $z = 0$, 
in excellent agreement with the GWTC-5.0 results. 
This is moderately higher than some previous predictions based on GC models with low binary fractions \citep[e.g.,][]{RodriguezLoeb2018, AntoniniGieles2020prd}, 
but close to the rate derived from the \catalog\ by \citetalias{Kremer2020catalog}.
We derive a redshift evolution parameter $\kappa \simeq 1.7$, at the low end of the GWTC-5.0 posterior distribution for {\sc Default BBH} \citep{LVK2026_populations}.
At $z > 1$, our rate prediction falls short of the {\sc Default BBH} result 
but remains consistent with {\sc PixelPop}. 
Meanwhile, our prediction naturally reproduces the observed merger rate at $z = 0.2$ in each mass bin under the {\sc Default BBH} inference. 
It also agrees with the {\sc Binned Gaussian Process} and {\sc PixelPop} (not shown) inferences in the two bins with $m_{1} < 50 \MSol$; 
however, both {\sc Binned Gaussian Process} and {\sc PixelPop} support merger rates up to an order of magnitude greater than {\sc Default BBH} at higher BH masses. 

At $z=0.2$, primordial binary, 1G dynamical, and hierarchical mergers 
are predicted to contribute approximately $9.6\%$, $85.4\%$, and $5.0\%$ of the total merger rate, respectively; 
the four mass bins described above contribute $76.9\%$, $22.2\%$, $0.8\%$, and $0.1\%$, in order of increasing mass. 
We find that different BBH formation channels and mass bins within dense clusters undergo differential rate evolution, 
meaning that their $\kappa$ values can differ significantly. 
Recalling Fig.~\ref{fig:masses_and_delays}, we can easily understand this as a byproduct of their different underlying delay-time distributions. 
1G dynamical mergers dominate the total rate by formation channel, and $m_{1} < 20 \MSol$ mergers by mass. 
These subsets have long delay times, 
resulting in shallow rate evolution with $\kappa = 1.6$ (1G dynamics) or $1.3$ ($m_{1} < 20 \MSol$). 
For comparison, using the \catalog, \citet{YeFishbach2024} find $\kappa \simeq 1$ across all mergers (dominated by 1G dynamics) and $\simeq 0.9$ for $m_{1} <20 \MSol$ mergers; 
\citet{AntoniniGieles2020prd} also predict $\kappa \simeq 1.6$ using semi-analytical methods.
On the other hand, both primordial binary evolution and $m_{1} > 20 \MSol$ mergers within dense clusters favor short delay times ($t_{\rm GW} \lesssim 1 \Gyr$). 
Accordingly, we find that their cosmic rate evolution closely traces the fiducial GC birth rate, 
with $\kappa \simeq a_{z} = 2.6$. 
This arises from two effects: 
First, clusters formed at higher redshifts have lower metallicities, thus producing larger BHs on average \citep{YeFishbach2024, Ye+2026}. 
Second, mass segregation within parent clusters 
causes more-massive BHs to undergo dynamical encounters more quickly, 
thereby forming BBHs with shorter delay times (see Fig.\ \ref{fig:masses_and_delays} and \citealt{YeFishbach2024}). 
Finally, for hierarchical mergers, we obtain $\kappa = 1.9$, between the values for 1G dynamics and primordial binary mergers. 
As noted in \S\ref{s:CMC:channels},
this deviates from previous predictions based on the \catalog\ \citepalias[e.g.,][]{Kremer2020catalog}, 
for which the 1G dynamics and hierarchical delay time distributions (and hence their redshift evolution) are nearly identical.
The finding that the 1G dynamics and hierarchical channels 
can undergo differential redshift evolution 
presents a complication for studies seeking to use hierarchical mergers 
as a tracer for the GC contribution to the LVK merger rate \citep[e.g.,][]{Farah+2026}. 

Before moving on, we consider the sensitivity of our predictions to different assumptions about the cosmic population of dense star clusters, 
particularly the magnitude and shape of the birth rate $\mathcal{R}_{\rm GC}(z)$. 
Since our merger rates are anchored to the local spatial density of GCs, 
they may be rescaled by an overall factor of $(n_{\rm GC}/2.1 \, {\rm Mpc}^{-3})$. 
Our fiducial value lies near the middle of the range of current observational estimates, 
which admit uncertainties of a factor of a few \citep[e.g.,][]{Harris+2013gccounts, DornanHarris2025}. 
The shape of the birth rate function $\mathcal{R}_{\rm GC}(z)$ is also open to plausible variations. 
We have fiducially assumed that dense star cluster formation traces the overall cosmic star formation history and hence peaks at $z_{\rm p} \simeq 2$ \citep[e.g.,][]{Elmegreen2010, Kruijssen2015, ReinaCampos+2022}. 
However, some theoretical models and observational assessments suggest that GC formation may have peaked earlier, 
with estimates stretching as far back as the pre-reionization era \citep[e.g.,][]{Forbes+2015, Forbes+2018, Trenti+2015, ElBadry2019gcformation, Ma+2021, Chisholm+2026}.

\begin{figure}
    \centering
    \includegraphics[width=\linewidth]{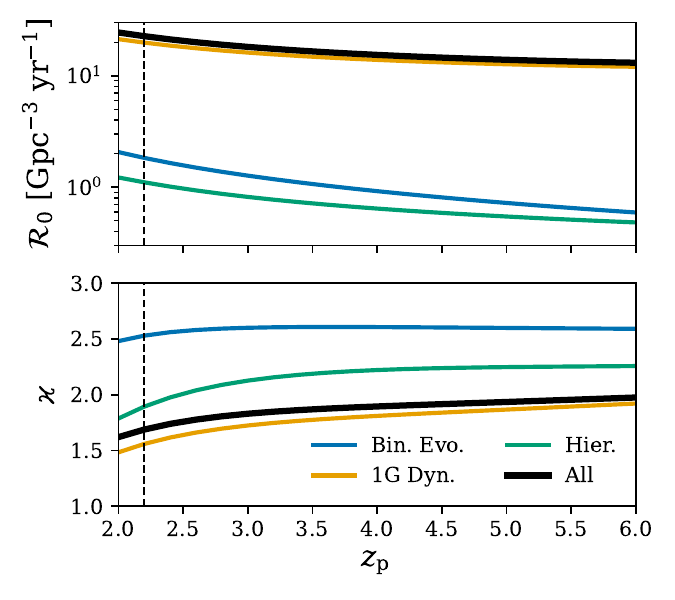}
    \caption{Cluster-derived BBH merger rate $\mathcal{R}_{0}$ and redshift evolution of parameter $\kappa$ 
    as functions of the redshift $z_{\rm p}$ at which the cosmic cluster formation rate density $\mathcal{R}_{\rm GC}$ peaks. 
    As in the left-hand panel of Fig.\ \ref{fig:rate_redshift}, the black curve indicates the total merger rate, 
    and the blue, amber, and green curves indicate the respective contributions of primordial binaries, 1G dynamics, and hierarchical mergers. 
    The parameters $n_{\rm GC}$, $a_{z}$, and $b_{z}$ are held fixed at their fiducial values. 
    The dashed vertical line marks $z_{\rm p} = 2.2$.}
    \label{fig:R0_kappa_zp}
\end{figure}

To gauge the impact of these uncertainties, we have repeated our rate calculations over a finely sampled grid $z_{\rm p} \in [2, 6]$, 
holding the shape parameters $a_{z}$ and $b_{z}$ and the cumulative density $n_{\rm GC}$ fixed. 
In Figure \ref{fig:R0_kappa_zp}, we plot the $\mathcal{R}_{0}$ and $\kappa$ values that we derive at $z \leq 1$ 
for the total merger rate and the three major in-cluster formation channels as functions of $z_{\rm p}$. 
In brief, $\mathcal{R}_{0}$ drops by a factor of $\sim 1.5 \mbox{--} 2$ between $z_{\rm p} = 2$ and $z_{\rm p} = 6$, depending on the channel in question. 
Meanwhile, the redshift evolution at $z < 1$ steepens with increasing $z_{\rm p}$. 
For primordial binaries and 1G dynamical mergers, $\kappa$ increases by a modest margin of $0.1 \mbox{--} 0.3$. 
For hierarchical mergers, the change is more dramatic, from $\kappa(z_{\rm p} = 2) \approx 1.8$ to $\kappa(z_{\rm p} = 6) \approx 2.5$. 
All of these changes occur because the bulk of cluster-derived mergers happen earlier, i.e., at higher redshifts, as $z_{\rm p}$ increases. 

\subsection{Mass and mass ratio distributions} \label{s:Results:m1_q}

\begin{figure*}
    \centering
    \includegraphics[width=\linewidth]{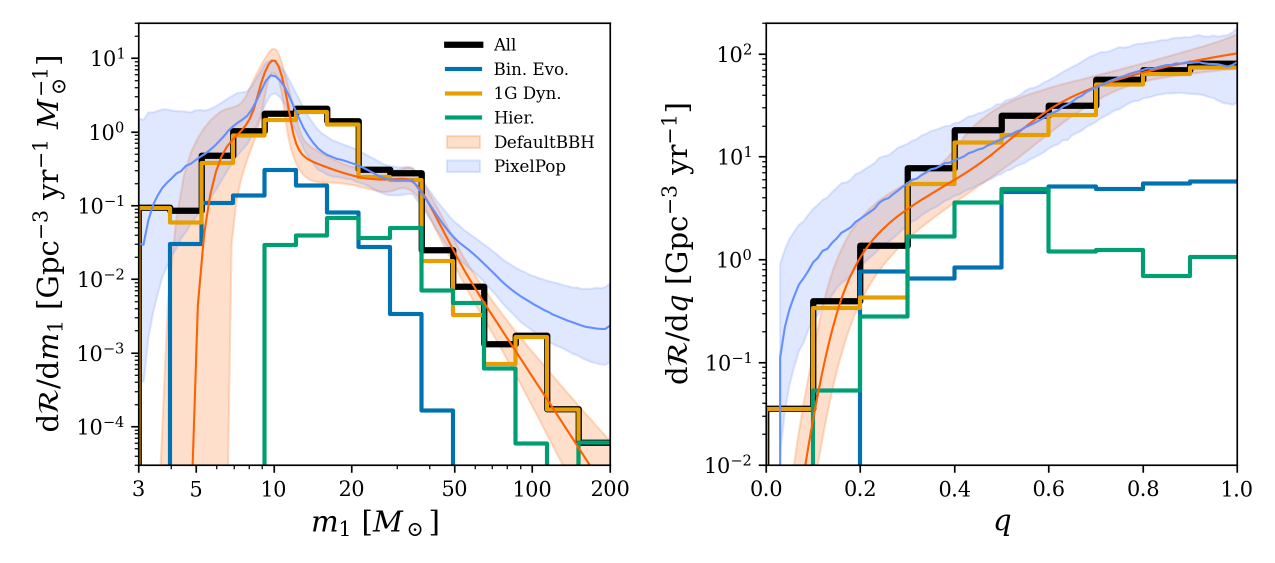}
    \caption{Comparison between the predicted differential rate of BBH mergers (histograms) with respect to the primary BH mass $m_{1}$ (left) and mass ratio $q$ (right). 
    The black histogram gives the combined merger rate over all formation channels, 
    while the blue, amber, and green give the contributions of primordial binaries, 
    1G dynamical assembly, and hierarchical mergers. 
    Each merger is weighted according to its parent model's cosmological contribution at $z = 0.2$. 
    The marginal posterior distributions are overplotted for the GWTC-5.0 {\sc Default BBH} (light red) and {\sc PixelPop} (light blue) models,
    with thin curves for medians and shading for 90\% credible regions.}
    \label{fig:distributions_m1q}
\end{figure*}

\begin{figure*}
    \centering
    \includegraphics[width=\linewidth]{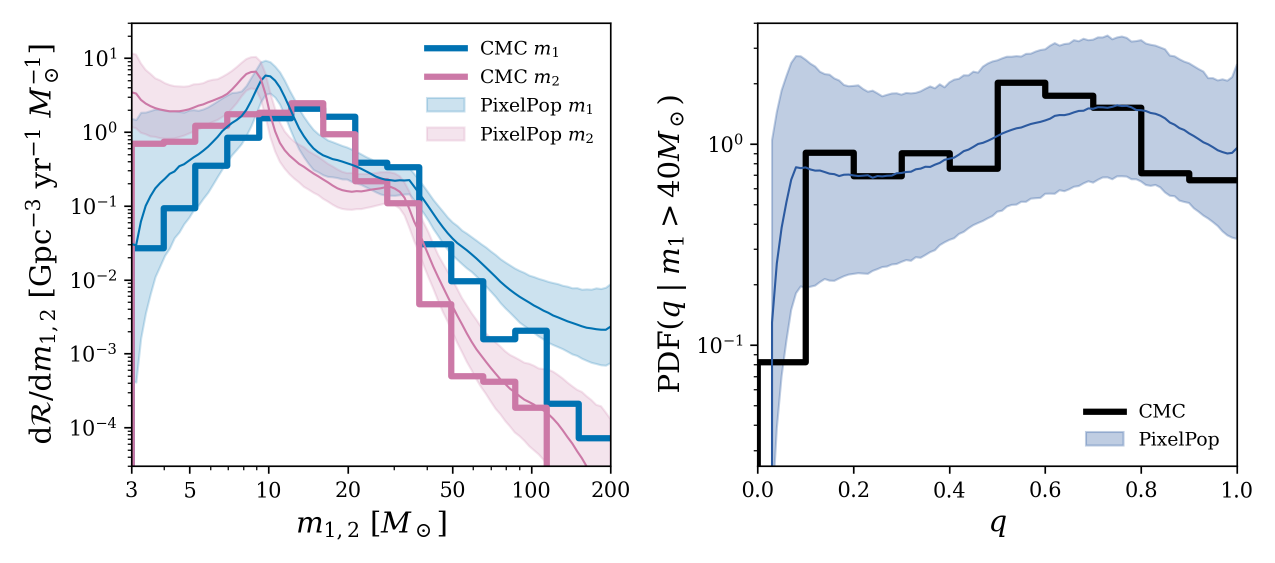}
    \caption{{\it Left:} Similar to the left panel of Fig.\ \ref{fig:distributions_m1q}, but now showing the distributions of the primary (blue) and secondary (purple) BH masses $m_{1,2}$ 
    for all mergers, regardless of formation channel. 
    {\it Right:} Similar to the right panel of Fig.\ \ref{fig:distributions_m1q}, 
    but now showing the mass-ratio distribution among all mergers with primary BH mass $m_{1} > 40 \MSol$. 
    Our \cmc-based prediction (black histogram) and the {\sc PixelPop} median posterior (thin, dark blue curve) are normalized to have unit area on the interval $0 < q \leq 1$.}
    \label{fig:dRdm12_pq}
\end{figure*}

The expected distribution function for merger properties at redshift $z$ 
may be estimated by assigning weights to each merger produced from our \cmc\ models 
according to the cosmological formation rate of parent clusters and the delay times of individual mergers \emph{\`{a} la} equation \eqref{eq:weighting_equation}. 
In Figure \ref{fig:distributions_m1q}, we show histograms approximating the predicted marginal distributions of the primary BH mass $m_{1}$ and mass ratio $q = m_{2}/m_{1}$ 
at redshift $z = 0.2$, expressed as differential merger rates $\dif \mathcal{R} / \dif m_{1}$ and $\dif \mathcal{R} / \dif q$, and normalized to a total rate of $\mathcal{R}(z=0.2) = 32.5 \, {\rm Gpc}^{-3} \yr^{-1}$. 
We also show the separate contribution of each BBH formation channel 
and superimpose the GWTC-5.0 posteriors under the {\sc Default BBH} and {\sc PixelPop} models.
The weighted medians and central 90\% quantile intervals of $m_{1}$ and $q$ are listed in Table \ref{tab:rates}.

Our prediction naturally captures the major qualitative features of the observationally inferred $m_{1}$ and $q$ distributions, 
similar to, but with certain differences from, predictions based on the \catalog\ \citep{Ye+2026}. 
Our predicted merger rate exhibits a peak at $m_{1} \simeq 8 \mbox{--} 20 \MSol$,
comparable to, but broader than, the prominent $\simeq 10 \MSol$ peak in the GWTC-5.0 posteriors. 
Interestingly, the \catalog\ produces a narrower peak in this regime, 
which better matches the observations at face value; 
this difference may be due to the higher binary fraction in our models, 
which facilitates the production of larger BHs through mergers between progenitor stars, 
or our use of the delayed SN explosion recipe as opposed to the rapid.
We further obtain a reasonable match to the location of the $35 \MSol$ feature
and to the slope of the distribution at high primary masses ($m_{1} \gtrsim 40 \MSol$).
However, we should note that the GWTC-5.0 results are highly model-dependent in the high-$m_{1}$ regime, 
where the data are sparser. 
Our prediction is in good agreement with the strongly modeled {\sc Default BBH} posterior distribution 
but falls definitely below the more weakly modeled {\sc PixelPop} result. 

Considering now the breakdown by formation channel, 
we find that the 1G dynamics channel dominates the total merger rate everywhere, except perhaps at $\simeq 50 \mbox{--} 75 \MSol$ 
(the lower part of the putative PISN mass gap), where hierarchical mergers are of comparable or greater frequency. 
Primordial binaries within dense clusters produce mostly lower-mass ($m_{1} < 20 \MSol$) mergers, 
with a steep decline in their contribution at higher masses and a sharp cutoff at $m_{1} \simeq 45 \MSol$. 
We tentatively attribute this to the prevalence of common-envelope evolution among these systems: 
the stable-mass-transfer channel produces wider and more massive BBHs, 
which are more susceptible to dynamical processes between formation and merger (see \S\ref{s:CMC:channels}). 
Hierarchical mergers occur across a wide range of masses in our models, starting at $m_{1} \simeq 10 \MSol$ and extending above $200 \MSol$, 
with power-law-like decline at $m_{1} \gtrsim 40 \MSol$; 
this compares favorably with recent analyses of LVK data targeting a hierarchical-merger component \citep{Farah+2026, Ray+2026}. 

Turning to the mass ratio $q$, we find excellent agreement between our fiducial prediction and the GWTC-5.0 posteriors. 
The merger rate increases monotonically towards $q = 1$ and drops off steeply below $q = 0.3$. 
This shape is naturally reproduced by dynamics in dense clusters 
because mass segregation strongly favors encounters between similar-mass BHs. 
Meanwhile, primordial binary evolution within clusters also moderately favors near-equal BH masses, but the distribution is flatter than that produced by dynamics. 
This is more or less consistent with the $q$ distribution produced by binary evolution in field populations \citep[e.g.,][]{Bavera+2021, BanerjeeOlejak2024}. 
Finally, hierarchical mergers preferentially have $q \simeq 0.5$ (as expected for 2G+1G pairings in mass-segregated cluster environments; e.g., \citealt{GerosaBerti2017, Rodriguez+2019, Kimball+2020, Ye+2026}), 
with tails extending up to $q \simeq 1$ and down to $q \simeq 0.1$. 

Analysis of GWTC-5.0 data reveals significant differences between the mass spectra of the primary ($m_{1}$)
and secondary ($m_{2}$) components of BBH mergers \citep{LVK2026_populations}. 
In particular, the incremental distribution of $m_{2}$ is flatter than that of $m_{1}$ below $10 \MSol$. 
Both distributions steepen towards high BH masses, 
but the change in slope occurs at different locations, $m_{1} \simeq 40 \MSol$ versus $m_{2} \simeq 30 \MSol$.
At the same time, the high-mass regime ($m_{1} \gtrsim 40 \MSol$) features a nearly flat distribution of $q$, 
contrasting with the steeply sloped distribution among lower-mass systems (Fig.\ \ref{fig:distributions_m1q}). 
Figure \ref{fig:dRdm12_pq} shows that our models naturally reproduce these characteristics, at least at a qualitative level.

\subsection{Spins} \label{s:Results:spins}

\begin{figure*}
    \centering
    \includegraphics[width=\linewidth]{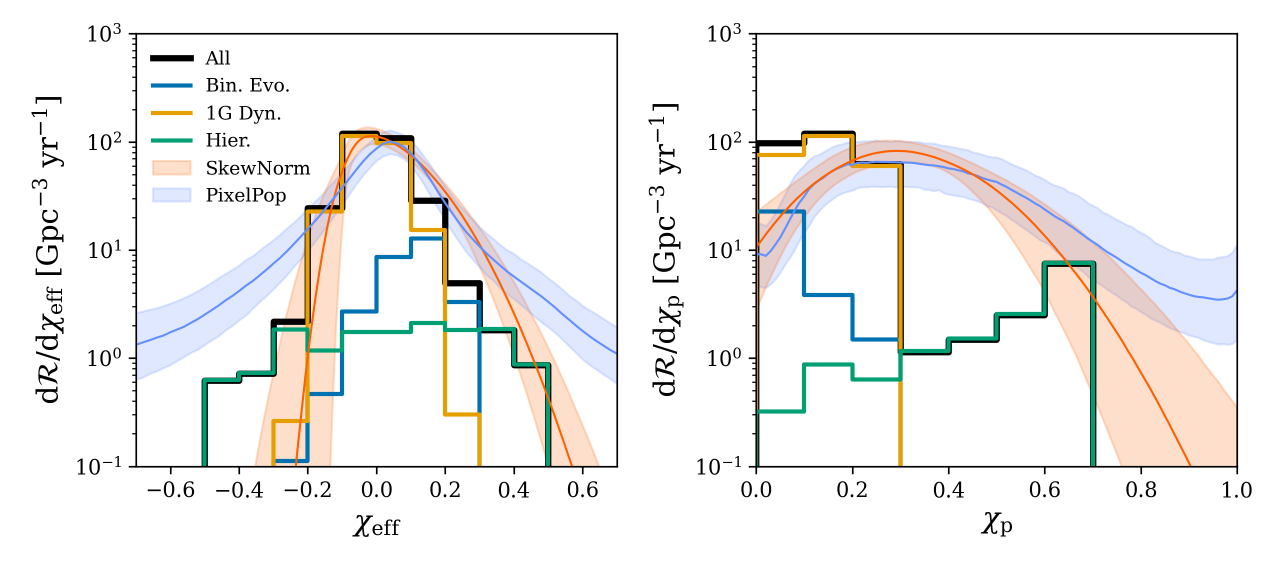}
    \caption{Similar to Fig.\ \ref{fig:distributions_m1q}, but for the effective inspiral spin $\chi_{\rm eff}$ (left) and effective precessing spin $\chi_{\rm p}$ (right). 
    GWTC-5.0 marginal posteriors are shown for the {\sc Bivariate Skewed} (light red) and {\sc PixelPop} (light blue) spin models.}
    \label{fig:distributions_spins}
\end{figure*}

In our \cmc\ simulations, all BHs are assumed to have zero spin at birth \citep[e.g.,][]{FullerMa2019}, 
and they can only acquire spin through mergers with other BHs. 
However, GW observations at this point confidently favor small, nonzero spins in most BBH mergers \citep{LVK2025_populations, LVK2026_populations}.
For the present, we offer a fiducial prediction of the effective spins of merging BHs based on post-processing (see \S\ref{s:Discuss:Caveats} for a discussion of the associated uncertainties). 

We assign each 1G BH a birth spin $0 \leq \chi \leq 0.3$ with uniform probability; 
in principle, the spins of 2G BHs should be modified to account for finite 1G spins, but we neglect this as a small correction \citep[see][]{Rodriguez+2019}. 
For each merger, we calculate the effective inspiral spin
\begin{equation}
    \chi_{\rm eff} = \frac{m_{1} \chi_{1}^{\parallel} + m_{2} \chi_{2}^{\parallel}}{m_{1} + m_{2}},
\end{equation}
and the effective precessing spin 
\begin{equation}
    \chi_{\rm p} = \max\left[ \chi_{1}^{\perp}, \frac{q (4q+3)}{3q+4} \chi_{2}^{\perp} \right].
\end{equation} 
The components of the spin of $m_{k}$ ($k = \{1,2\}$, $m_{1} \geq m_{2}$) parallel and perpendicular to the orbital angular momentum are given by 
\begin{equation}
    \chi_{k}^{\parallel} = \chi_{k} \cos{\theta_{k}}, \hspace{0.25cm}
    \chi_{k}^{\perp} = \chi_{k} \sin{\theta_{k}}, 
\end{equation}
where $\theta_{k}$ is the tilt angle between the unit vector $\lhat$ along the orbital angular momentum of the merging binary 
and the unit vector $\vect{\hat{s}}_{k}$ along the spin of $m_{k}$.

The methods by which we sample the spin--orbit orientations, thereby obtaining $\chi_{\rm eff}$ and $\chi_{\rm p}$, differ for each merger channel we have identified.
For dynamically formed mergers, we assume that the spin directions are uncorrelated and isotropically distributed with respect to $\lhat$. 
Thus, to sample the $\chi_{\rm eff}$ distribution we draw $\theta_{1}$ and $\theta_{2}$ independently and identically 
from the spherical isotropic distribution $p(\cos\theta) = {\rm const.}$, $\cos\theta \in [-1,1]$. 

Meanwhile, for the quasi-isolated binary evolution channel, 
we assume that BHs have a strong intrinsic preference for spin--orbit alignment at birth 
due to tidal synchronization and/or mass transfer in close binaries \citep[e.g.,][]{FullerMa2019, Kiroglu2025spinorbitalign}, 
so that $\vect{\hat{s}}_{k} = \lhat$, where the right-hand side is evaluated \emph{just before} BH formation. 
However, we allow SN kicks to reorient $\lhat$ prior to merger. 
Because \cmc\ does not save kick-direction data from \cosmic\ at present, 
we sampled the spin--orbit tilt angles from our auxiliary \cosmic\ dataset (see \S\ref{s:CMC:Methods:COSMICauxiliary}). 
For each quasi-isolated BBH in \cmc\ with component masses $m_{1}$ and birth metallicity $Z$, we adopt the SN kick data of the most similar \cosmic\ BBH,
as determined by minimizing the quantity
\begin{align}
    \delta_{l}^{2} = \left( \frac{m_{1,l}}{m_{1}} - 1 \right)^{2} + \left( \frac{m_{2,l}}{m_{2}} - 1 \right)^{2} + \left( \frac{Z_{l}}{Z} - 1 \right)^{2},
\end{align}
where $l$ is the index for mergers in our \cosmic\ dataset. 
This does not guarantee that the evolutionary histories of the individual systems are similar, 
but since our object here is only a rough characterization of the spin statistics associated with in-cluster binary evolution, 
perfect consistency is not required. 
We have verified that sampling these quantities at random from the \cosmic\ dataset yields qualitatively similar results. 

In principle, mergers from the perturbed binary evolution channel should begin with spin--orbit tilts similar to those of the quasi-isolated BBHs 
and have their tilts modified in the course of in-cluster encounters. 
Qualitatively, we expect the tilt distribution for perturbed binaries to depend on the total number of encounters $N_{\rm enc}$ 
and the average mass ratio between the `target' binary and the additional bodies it encounters. 
Moreover, it should resemble the isolated-binary tilt distribution in the limit $N_{\rm enc} \to 0$ 
and converge on an isotropic distribution as $N_{\rm enc} \to \infty$. 
However, the intermediate behavior is complicated. 
A forthcoming parallel work (M.\ A.\ S.\ Martinez et al., in prep.) will present a rigorous statistical characterization of the cumulative effect of serial in-cluster encounters on a target binary, 
subject to the constraint that the target survive as a bound pair. 
For now, we assume that perturbed BBHs achieve isotropically distributed spins by the time they merge, like the dynamically assembled systems, regardless of the number and character of previous encounters. 

In Figure \ref{fig:distributions_spins}, we compare the predicted distributions of $\chi_{\rm eff}$ and $\chi_{\rm p}$, again as evaluated at $z = 0.2$, 
with results from GWTC-5.0 under the {\sc Bivariate Skew-Normal} and {\sc PixelPop} models. 
Considering $\chi_{\rm eff}$ first, both GWTC-5.0 models provide strong support for a broad distribution of effective spins 
with moderate asymmetry about $\chi_{\rm eff} = 0$. 
Our fiducial prediction is in good agreement with both models for $|\chi_{\rm eff}| \lesssim 0.2$. 
Extending the distribution to more extreme spins 
would require that we consider birth spins above $0.3$, 
for which feedback effects on cluster dynamics cannot be neglected (see \S\ref{s:Discuss:Caveats}).

It is of interest to note that different BBH formation channels operating within the same population of clusters 
produce markedly different $\chi_{\rm eff}$ distributions.  
The 1G dynamics channel produces a Gaussian-like distribution centered at $\chi_{\rm eff} = 0$ with a $90\%$ spread of $\pm 0.11$. 
Systems derived from in-cluster binary evolution prefer positive $\chi_{\rm eff}$ ($81\%$ have $\chi_{\rm eff} > 0$) but a definite minority has $\chi_{\rm eff} < 0$, 
reflecting orbital realignment of primordially paired BBHs by a combination of SN kicks and dynamical interactions.  
Finally, hierarchical mergers have a symmetric and nearly flat distribution, roughly occupying the interval $-0.5 \lesssim \chi_{\rm eff} \lesssim 0.5$. 

Turning now to the precessing spin parameter $\chi_{\rm p}$, 
we find greater discrepancies than before between our fiducial prediction and the constraints derived from GWTC-5.0. 
As before, the 1G dynamical channel dominates the shape of the distribution for $\chi_{\rm p} \leq 0.3$. 
Hierarchical mergers make their presence known at higher values, 
producing a secondary peak at $\simeq 0.7$ \citep{GerosaBerti2017, Rodriguez+2019}. 
Broadly speaking, we produce widely distributed $\chi_{\rm p}$ as in the observations below $\chi_{\rm p} = 0.3$, but we significantly under-predict the rate of mergers with higher $\chi_{\rm p}$.  
In other words, our fiducial treatment of BH spins under-predicts the \emph{in-plane} components of BH spins. 
Earlier analyses of BBH demographics through GWTC-4.0 have noted the difficulty 
of matching both the $\chi_{\rm eff}$ and $\chi_{\rm p}$ posterior distributions under a purely isotropic spin distribution \citep[e.g.,][]{LVK2025_populations}. 
Multiple interpretations of this fact have been proposed, 
including secular dynamics in hierarchical triples as a dominant BBH formation channel \citep{Stegmann+2026} 
and spin realignment through accretion during BH--star (or even BBH--star) collisions in dense clusters \citep{Kiroglu2025spinup, Kiroglu2025spinorbitalign}.

\section{Discussion} \label{s:Discussion}

\subsection{\cmc\ versus GWTC}

We have found that dense star clusters are capable of producing a large fraction 
of the BBHs observed as GW sources at low redshift \citep[see also][]{Rodriguez+2021}. 
Our results elaborate on previous studies of BBH formation within clusters by implementing a realistic, 
observationally motivated model of the initial binary population based on field stellar populations. 
However, there are a few aspects of the inferred demographics of BBHs that our models struggle to reproduce under simplistic assumptions. 

The most important difference is the location and width of the main peak in the BH mass spectrum: 
GW observations strongly favor a fairly narrow peak at $\simeq 10 \MSol$, 
but our models produce a broader peak at $\simeq 15 \MSol$. 
Various explanations for the $10 \MSol$ feature have been proposed within the isolated-binary-evolution paradigm, 
such as failed SN explosions \citep{DisbergNelemans2023, Legred+2026} and stripped stellar evolution  \citep{Schneider+2023}. 
It is not obvious how these elaborations of stellar evolution would affect the mass spectrum of dynamically assembled BBHs (but see \citealt{Galaudage2026} for an analysis of GWTC-4.0 through this lens).
It is interesting to note, however, that the subset of mergers produced by primordial binary evolution in our models 
does exhibit a peak at $m_{1} \simeq 10 \MSol$ even under our fiducial binary evolution prescriptions. 

The second point of tension is the under-production of mergers with high in-plane spin components (Fig.\ \ref{fig:distributions_spins}). 
\citet{Stegmann+2026} have argued in favor of a formation channel that yields spin--orbit tilt angles clustered about $90^{\circ}$, 
such as secular dynamics in isolated hierarchical triple systems. 
However, we remark that an isotropic distribution of spin tilts is, and has always been, 
an \emph{assumption} made by theoretical studies of BBH formation via cluster dynamics. 
Recent studies have identified accretion during BH--star and BBH--star collisions 
as a potential path to spin--orbit alignment in dynamically active cluster environments \citep{Kiroglu2025spinup, Kiroglu2025spinorbitalign}. 
Additionally, future cluster models may be able to track BH spins as part of dynamical encounters 
by building on post-Newtonian methods for direct integrations \citep[e.g.,][]{Rodriguez2018cmc2.5pn, Rodriguez2018cmc2.5pn2}, 
allowing a bona fide \emph{prediction} of the dynamically generated spin--orbit tilt distribution in gas-free environments.

\subsection{Cluster properties}

In keeping with our focus on BBH mergers in this paper, 
we have eschewed a detailed discussion of the global dynamical evolution of our models. 
However, some comparisons with real YMCs and GCs are necessary to demonstrate 
that our model grid is, at a basic level, plausibly representative of a cosmological population of dense clusters 
(see, e.g., \citetalias{Kremer2020catalog}). 
Since our initial cluster profile and binary population model are based in large part on observations of nearby YMCs
(see \citealt{ProtegiesZwart2010ymcreview}, \citealt{Offner2023binaries}, and references therein), 
we need not recapitulate these details.
We make two points regarding our models' late-time ($t \gtrsim 10 \Gyr$) properties. 

First, about half of our simulations reach a core-collapsed state within a Hubble time. 
This matters in light of the connection between the onset of core collapse and the depletion of a GC's central BH subsystem: 
core-collapsed clusters are thought to have ``burned'' nearly all their original BHs through the dynamical formation and ejection of hard BBHs, 
thereby exhausting the energy source that supported their cores \citep{BreenHeggie2013, Kremer2019corecollapse, Kremer2020bhburning}. 
Second, our models exhibit total (core) binary fractions of $\simeq 5 \mbox{--} 15 \%$ ($\simeq 10 \mbox{--} 30\%$) at late times, 
similar to the range observed among old Galactic GCs \citep[e.g.,][]{Milone2012gcbinaryfraction, Ji2015_gcbinaries}. 
In that sense, our models capture the diversity of dynamical states and stellar demographics of both old Galactic GCs 
\emph{and} resolved MS populations in YMCs and the Solar neighborhood. 

\subsection{Field vs.\ cluster environments}

As mentioned in the Introduction, the fact that star formation occurs predominantly in clusters \citep{LadaLada2003}
prompts consideration of what separates the ``field'' and ``cluster'' stellar populations as distinct formation environments for BBHs. 
Our simulations, by subjecting a realistic, evolving population of massive stellar binaries to in-cluster dynamical perturbations, 
reveal that massive binaries can undergo effectively isolated evolution into merging BBHs within dense clusters, 
provided that the isolated delay time be less than the parent cluster's mass-segregation time (Fig.\ \ref{fig:masses_and_delays}). 

The smallest cluster that can, in principle, produce a BBH is 
one that contains 2 stars above $\simeq 20 \MSol$; under a canonical Kroupa IMF, 
such a cluster contains $N \simeq 2000$ stars in total (or a stellar mass $M_{0} \simeq 1200 \MSol$). 
By equations \eqref{eq:Trlx_def} and \eqref{eq:Tseg_def},
the relaxation and mass-segregation timescales of that same cluster are $T_{\rm rlx} \simeq 10 \Myr$ and $T_{\rm seg} \simeq 1 \Myr$.
Fortuitously, these characteristic values are comparable to the typical lifespans of O-type stars; 
likewise, the shortest delay times realized among quasi-isolated binaries in our \cmc\ simulations were $t_{\rm GW} \simeq 4 \Myr$. 
Thus, a massive binary within any cluster that remains bound for more than a few Myr 
is likely to experience significant dynamical processing before it merges as a BBH. 
This comports with our finding that primordial binary evolution contributes no more than $\simeq 10\%$ of the merger rate derived from massive clusters (\S\ref{s:Results:RatesChannelsMasses}) in the local Universe.

It follows from the above that the ``field'' component of the BBH population properly corresponds to BH progenitors born in clusters with lifetimes $\ll T_{\rm rlx}$, 
such as those that disperse promptly following gas expulsion. 
The ``cluster'' component corresponds to those whose parent clusters remained bound for a time $\gtrsim T_{\rm rlx}$. 
However, since unbound clusters predominantly have birth masses $\lesssim 10^{3} \MSol$ \citep{LadaLada2003}, 
there may be only a narrow range of birth masses in which clusters are simultaneously rich enough 
to contain multiple BH progenitors and tenuous enough to disperse rapidly. 
We conclude that efforts to characterize the observed BBH population 
as a mixture of field- and cluster-derived subpopulations \citep[e.g.,][]{ArcaSedda+2026, Ray+2026, Galaudage2026} 
may need to account for the clustered nature of massive star formation and the nuances of birth cluster dispersal. 


\subsection{Related and future works}

We have already made a number of comparisons between the results of this work 
and those based on the \catalog\ (\citetalias{Kremer2020catalog}; \citealt{FishbachFragione2023}; \citealt{YeFishbach2024}; \citealt{Ye+2026}). 
To briefly recap, we have found moderately different rates of cluster-derived BH mergers, featuring notably steeper redshift evolution at $z < 1$, 
as well as moderately different BH demographics, such as a somewhat larger typical mass. 
We have described how these changes are variously linked to the higher binary fraction in our models, 
the gist being that binaries accelerate the formation and dynamical processing of BBHs within clusters. 

Our study likewise builds on that of \citet{Hong2018bbhmergers}, 
who used the {\tt MOCCA} code to explore BBH formation across the space of GC parameters, including initial binary fraction. 
That work also identified primordial binary evolution and dynamical formation as distinct BBH formation channels within dense clusters. 
In addition to using a realistic, mass-dependent initial binary population,
we have gone into greater depth by distinguishing between quasi-isolated and perturbed binaries, 
as well as between 1G and hierarchical dynamical mergers. 

Multiple direct $N$-body GC model grids with initial binary populations similar to ours are reported to be underway, 
including {\tt DRAGON-III} \citep{Wu+2025dragon3} and {\tt TITANS} \citep{Mestichelli2026imbhs}. 
As of this writing, published results from these projects relevant to this work pertain to the formation of IMBHs at early times ($< 100 \Myr$). 
\citet{Mestichelli2026imbhs} report that the formation of IMBHs with mass $> 140 \MSol$ via chained stellar collisions is inefficient in the {\tt TITANS} models, 
with hierarchical mergers dominating the formation of such objects; 
in contrast, our models have produced IMBHs up to $\simeq 400 \MSol$ via stellar collisions, 
with up to 4 in a single simulation (see Table \ref{tab:bbhraw_v2} and Fig.\ \ref{fig:masses_and_delays}). 
The difference may simply come down to cluster density: the {\tt TITANS} grid mostly probes clusters with lower densities than ours. 
Given that collisional evolution is highly sensitive to cluster density \citep{Kremer2020massgap, GonzalezPrieto+2021, GonzalezPrieto+2024}, 
the greater yield of IMBHs from our \cmc\ grid is not especially surprising. 
Meanwhile, \citet{Wu+2025dragon3} report the formation of a $(127 + 89) \MSol$ BBH within the first 100 Myr of a {\tt DRAGON-III} model with $N = 10^{6}$ and $r_{\rm v} \simeq 1.3 \pc$ (half-mass radius $r_{\rm h} = 1.75 \pc$). 
The left panel of Fig.\ \ref{fig:dRdm12_pq} shows that our simulations produce BBH mergers with similar pairings. 
On the whole, the relatively efficient formation of IMBHs in our \cmc\ models 
is plausibly consistent with these direct $N$-body results.
Further comparisons will no doubt prove illuminating when the full simulation results of these projects become available. 

The cluster model grid presented in this paper is a precursor to version 2 of the \catalog\
in that it reflects many (but not all) of the updates to \cmc\ and \cosmic\ that will be included in those models and explores a relatively small portion of parameter space. 
Yet to come are a prescription for BH accretion, spin-up, and spin realignment during failed common-envelope events and BH--star collisions \citep{Kiroglu2025spinup,Kiroglu2025bhaccretion,Kiroglu2025spinorbitalign}, 
updates to the geometric criteria that trigger {\tt FEWBODY} integrations (M.\ A.\ S.\ Martinez et al., in prep.),
and improved estimates of GW recoil kicks following BH mergers \citep[e.g.,][]{IslamWadekar2026, Ravichandran+2026}. 
Each of these changes may further modify the properties of BBH mergers produced in dense clusters, 
particularly those with extreme masses, mass ratios, or spins. 
Additionally, a concurrent suite of new \cmc\ models (S. Agrawal et al. 2026, in prep.) significantly refines the coverage in stellar metallicity relative to the original \catalog\ and the models presented here. 
That study demonstrates that cluster metallicity \citep[spanning the full range of observed values, including both ``blue" and ``red" cluster subpopulations; e.g.,][]{BrodieStrader2006,Peng2006,Strader2011} has an important effect on BBH merger demographics and large-scale cluster dynamics.
The next version of the \cmc\ catalog will also have at least as much parameter-space coverage as the first, also varying aspects not explored here such as different cluster orbits in the galaxy. 
We reiterate that our models provide representative predictions about the \emph{majority} of BBH mergers formed in the \emph{majority} of dense clusters throughout cosmic history. 

\subsection{Caveats} \label{s:Discuss:Caveats}

The initial condition for our cluster simulations, a spherically symmetric, non-mass-segregated King profile in virial equilibrium, represents a standard choice for $N$-body star cluster models. 
However, this choice omits certain factors potentially affecting the dynamics of YMCs, 
such as primordial mass segregation \citep[e.g.,][]{Vesperini+2009, Ghasemi+2024} and bulk rotation \citep[e.g.,][]{Bissenkov+2025, Bianchini+2026}. 
The absence of these effects in our models may mean that 
we underestimate the rate of scattering and collisions between massive stars at early times. 
This may have complex downstream consequences for BH formation and dynamics, as well as long-term cluster evolution.

Given our emphasis on realistic \emph{initial} binary properties in this work, 
we would be remiss to ignore the limitations of our approach to binary \emph{evolution}. 
As previously stated, there are mismatches between the outputs of rapid population synthesis codes 
and detailed binary evolution models regarding BBH formation specifically \citep{GallegosGarcia2021bbh}. 
Efforts to ameliorate these issues by, for instance, replacing or supplementing analytical prescriptions 
with interpolations over detailed single and binary stellar model grids 
\citep{Agrawal+2020, Agrawal+2025, Iorio+2023, Fragos+2023, Andrews+2025} are underway and advancing rapidly. 
Future improvements to the treatment of binary evolution within \cmc\ will be crucial 
to the refinement of predictions about the cluster-derived BBH population. 
Despite these uncertainties, we are confident in our conclusion that 
primordial binary evolution contributes a small fraction of the overall cluster-derived merger rate, 
since this is dictated by dynamical selection effects (see \S\ref{s:CMC:channels}). 

Our approach to BH spins in this work has been to inject 1G birth spins up to $\chi = 0.3$ in post-processing, 
neglecting finite-spin effects on BH dynamics within clusters. 
This is adequate for the primordial binary and 1G dynamics channels, but we should note the downstream implications for hierarchical mergers. 
When 1G BHs have zero spin, up to $\simeq 60\%$ of 2G BHs can be retained within typical GCs;
however, birth spins of $\chi \simeq 0.1 \mbox{--} 0.3$ can increase the typical GW recoil kick on a merger product, 
reducing the retention fraction by a factor of $\simeq 3 \mbox{--} 5$ \citep{Rodriguez+2019}. 
Consequently, by neglecting this effect, we may overestimate the rate of hierarchical mergers by a similar factor.
On the other hand, \citet{Islam+2026gwrecoil} have recently argued that longstanding analytical recipes to estimate GW recoil kicks, including that implemented in \cmc, 
can overestimate kick magnitudes in some regions of parameter space relevant to GC dynamics \citep[see also][]{IslamWadekar2026}. 
We plan to investigate the impact of different GW recoil recipes on cluster-derived hierarchical merger rates in future work. 
For now, we remark that our hierarchical merger rates should be viewed as upper bounds. 

\section{Summary} \label{s:Summary}

In this paper, we have used \cmc\ to study the formation and properties of merging binary black holes  
within dense star clusters with realistic initial binary populations 
over a range of masses, densities, and metallicities representative of old GCs and YMCs analogous to GC progenitors. 

Our main findings are as follows:
\begin{enumerate}
    \item Dense star clusters with high massive binary fractions produce BBH mergers 
    through a combination of primordial binary evolution and dynamical processes. 
    We have characterized the number of BBHs produced by the primordial binary evolution, 1G dynamical, and hierarchical merger channels 
    as functions of the parent cluster's mass, radius, and metallicity, 
    as well as the different delay time distributions associated with each channel. 
    At $z = 0.2$, $\simeq 85\%$ of mergers are dynamically assembled 1G systems, 
    $\simeq 10 \%$ are from primordial binaries, and $\simeq 5\%$ are hierarchical. 
    These estimates are subject to moderate systematic uncertainties related to 
    the treatment of binary stellar evolution and the effects of 1G BH spins.

    \item Using a fiducial model of a population of dense star clusters forming and evolving over cosmic time, 
    we find that the rate of cluster-derived BBH mergers and its redshift evolution in the local universe 
    are consistent with the total merger rate inferred from GW observations. 
    These results are mildly sensitive to the epoch at which the dense star cluster formation rate peaks. 

    \item Our models predict that the redshift evolution of the rate of hierarchical mergers formed within dense clusters 
    is significantly steeper at $z < 1$ than that of 1G dynamical mergers. 
    This contrasts with the prediction of previous model grids with low primordial binary fractions, such as the \catalog\ \citep[but see][]{Mai+2026},
    but is consistent with some recent analyses of GW source demographics \citep[e.g.,][]{Farah+2026, Ray+2026}. 

    \item Our models broadly reproduce the observationally inferred mass and mass-ratio distributions among BBH mergers. 
    Our predicted mass distribution includes key features such a peak at $\simeq 8 \mbox{--} 20 \MSol$, a break at $\simeq 35 \MSol$, 
    and differences in shape between the primary and secondary BH distributions. 
    The mass-ratio distribution is sharply peaked at $q \simeq 1$ overall 
    but nearly flat for primary BHs above $40 \MSol$, consistent with GWTC-5.0 data.
    A more sophisticated treatment of BH spins in \cmc, 
    encompassing 1G birth spins, subsequent spin-up physics, and on-the-fly spin dynamics during few-body encounters, 
    is required to match the observed $\chi_{\rm eff}$ and $\chi_{\rm p}$ distributions. 

    \item We find that IMBHs ($> 120 \MSol$) can form as a result of chained stellar collisions in massive ($N \gtrsim 4 \times 10^{5}$), compact ($r_{\rm v} \lesssim 1 \pc$) star clusters with metallicities $Z/Z_\odot \in [0.03, 1]$. 
    The predicted rate of BBH mergers involving these objects in the local universe
    is comparable to that inferred from LVK observations as of GWTC-5.0.
\end{enumerate}

\begin{acknowledgments}
    We thank Anarya Ray for helpful discussions and for providing the GWTC-5.0 \textsc{Binned Gaussian Process} inference data shown in Fig.\ \ref{fig:rate_redshift}. 
    We thank Dany Atallah, Carl Rodriguez, Salvatore Vitale, Newlin Weatherford, Noah Wolfe, and Mike Zevin for additional helpful discussions. 
    C.E.O.\ and F.K.\ acknowledge support from CIERA Postdoctoral Fellowships.
    Support for E.G.P.\ was provided by the NSF Graduate Research Fellowship Program under grant DGE-2234667. C.S.Y. acknowledges support from the Alfred P. Sloan Foundation.
    This work was supported by NSF Grants AST-2108624 and AST-2511543 at Northwestern University. 
    This work used computing resources at the Quest high-performance computing facility provided by CIERA under NSF Grant PHY-2406802.
    Quest is jointly supported by Northwestern University's Office of the Provost, the Office for Research, and Northwestern University Information Technology.

    \software{{\tt Astropy} \citep{Astropy2022}, {\tt Cluster Monte Carlo} \citep{Joshi2000, Pattabiraman2013, Rodriguez2022cmcreview}, {\tt COSMIC} \citep{Breivik2020cosmic}, {\tt Matplotlib} \citep{Hunter2007_matplotlib}, {\tt NumPy} \citep{Harris+2020_NumPy}, {\tt SciPy} \citep{Virtanen2020_scipy}}
\end{acknowledgments}

\bibliography{refs}
\bibliographystyle{aasjournalv7}

\end{document}